\documentclass[aps,prd,twocolumn,showpacs]{revtex4}

\usepackage{graphicx}

\usepackage{subfigure}

\usepackage{bbm}

\newcommand{\nn}{\nonumber\\}

\begin{document}

\title{Quasiconformality and mass}

\author{Dennis D.~Dietrich}

\affiliation{CP$^3$-Origins, Centre for Particle Physics Phenomenology, University of Southern Denmark, Odense, Denmark}

\date{May 5, 2010}

\begin{abstract}
We identify universal quasiconformal (walking) behaviour in non-Abelian gauge field theories based on the mass-dependent all-order $\beta$-function introduced in \cite{Dietrich:2009ns}. We find different types of walking behaviour in the presence of (partially) massive species. We employ our findings to the construction of candidate theories for dynamical electroweak symmetry breaking by walking technicolour.
\end{abstract}


\maketitle


\section{Introduction}

Here, we analyse further implications from the mass-dependent $\beta$-function introduced in \cite{Dietrich:2009ns}.

Regarding a variable number of active flavours breaks the two-loop scheme independence (``universality") of the $\beta$-function of non-Abelian gauge field theories. Considering a fixed number of species is legitimate for energy scales far above or far below their masses. From the viewpoint of renormalisation theory, it is even legal to ignore the freezing out of heavy flavours, although this disregards important physical effects and leads to a pure convergence of the perturbative series. Oftentimes $\beta$-functions for different integer numbers of flavours are glued together at the mass of the species which are switched on or off. Already with this procedure the $\beta$-function coefficients are no longer scheme independent in the above sense. The passage between different numbers of active flavours should happen gradually. Hence, we based the $\beta$-function in \cite{Dietrich:2009ns} on a background field momentum subtraction scheme \cite{Jegerlehner:1998zg}, which next to the decoupling theorem \cite{Appelquist:1974tg} also respects the Slavnov--Taylor identities \cite{DeWitt:1980jv,De Rujula:1976au,Georgi:1976ve,Nachtmann:1978vf,Shirkov:1992pc,Chyla:1995fm}. An alternative would have been the physical charge approach \cite{Brodsky:1999fr}, which, however, coincides to lowest order and is otherwise qualitatively and also quantitatively close to the background field momentum subtraction scheme (See especially Fig.~2 in \cite{Brodsky:1999fr}.) where the expressions are known analytically. It turns out that threshold effects are felt more then two orders of magnitude away from the mass of the fermion. In \cite{Dietrich:2009ns}, we combined this input with the all-order $\beta$-function without threshold effects that had been conjectured in \cite{Ryttov:2007cx}. It was motivated by the $\mathcal{N}$=1 supersymmetric Novikov--Shifman--Vainshtein--Zakharov $\beta$-function, has the correct limits in exactly known cases (super Yang--Mills theory \cite{Novikov:1983uc} or planar equivalence in a large-$N_c$ limit \cite{Armoni:2003gp}), and at two-loop order it reproduces the universal $\beta$-function. Far away from all thresholds our $\beta$-function reduces to the massless all-order $\beta$-function with a fixed number of flavours, while at two-loop order it coincides with the $\beta$-function in background field momentum subtraction scheme \cite{Jegerlehner:1998zg}.

\begin{figure}[t]
 \resizebox{8.5cm}{!}{\includegraphics{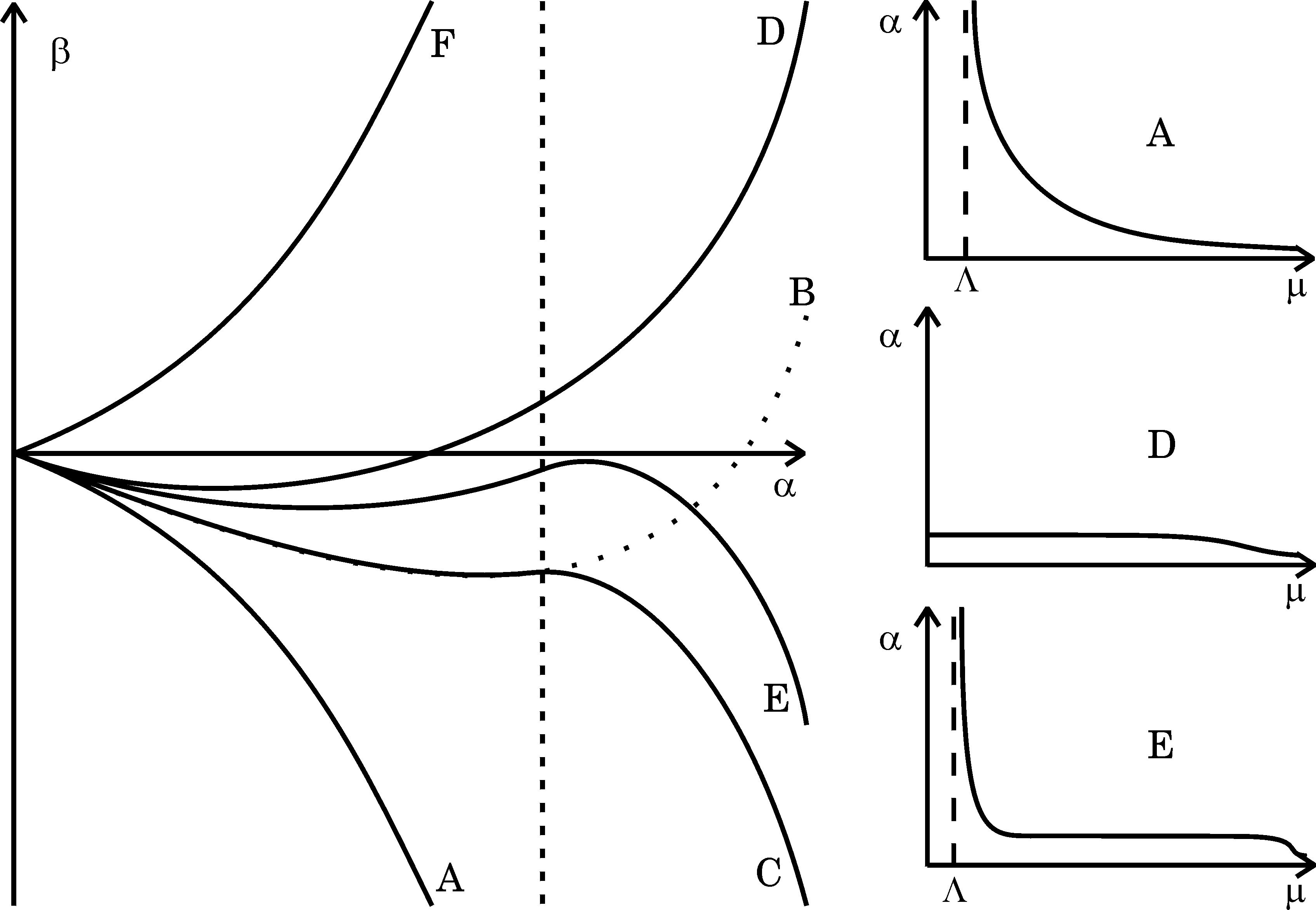}}
 \caption{Behaviour of the $\beta$-function as a function of the coupling $\alpha=g^2/(4\pi)$ and of the coupling as a function of the energy scale $\mu$, depending on the matter content of the theory.
 A) No or little matter;
 B) existence of a perturbative Caswell--Banks--Zaks fixed point;
 C) actual shape due to chiral symmetry breaking;
 D) realised fixed point;
 E) quasiconformal case;
 F) loss of asymptotic freedom.
 The dashed line in the plot on the left-hand side indicates the critical values of the coupling for chiral symmetry breaking. (Taken from \cite{Dietrich:2009ns}.)}
 \label{beta}
 \end{figure}

The mass-dependence of the $\beta$-function is important in the context of the conformal window of non-Abelian gauge field theories. It arises from the interplay of the matter content of a theory and chiral symmetry breaking (See Fig.~\ref{beta}.): With no or little matter [like in quantum chromodynamics (QCD)] the antiscreening of the non-Abelian gauge bosons forestalls the appearance of an infrared fixed point (A). Slightly more matter admits a perturbative Caswell--Banks--Zaks \cite{Caswell:1974gg} fixed point (B). For this fixed point to be reached it must be situated at a value of the coupling that does not trigger chiral condensation (D). [With even more matter the theory loses asymptotic freedom (F).] Otherwise the fermions receive a dynamical mass and decouple at least partially, which makes the antiscreening dominate once more (C). Where the fixed point is almost reached (E) but chiral condensation still sets in we find the quasiconformal case. In the vicinity of the would-be fixed point $\beta\lesssim 0$. This results in a coupling that stays almost constant (``walks") for a large interval of energy scales at a value slightly below the critical value for chiral condensation. Once chiral condensation is finally triggered the coupling constant begins running again.
What has just been said explains why taking into account the mass of the fermions is a crucial ingredient for gaining an understanding of the dynamics of the theory. 

One of our motivations for conducting research into the conformal window of gauge field theories and for the study \cite{Dietrich:2006cm} is the identification of quasiconformal \cite{walk} technicolour \cite{TC} models that are consistent with presently available electroweak precision data \cite{Dietrich:2005jn,Dietrich:2005wk}. These models feature a rich collider phenomenology accessible to the Large Hadron Collider (LHC) \cite{Belyaev:2008yj}, dark matter candidates \cite{Gudnason:2006ug}, and are interesting models for studies in the AdS/CFT framework \cite{Hong:2006si}. They break the electroweak symmetry dynamically by chiral symmetry breaking among fermions (techniquarks) added to the standard model without its Higgs sector. This gives masses to the weak gauge bosons. The canonical way to also give masses to the standard model fermions is extended technicolour (ETC) \cite{ETC}. In this context, the walking serves to relax the tension between the large mass of the top quark, which has to be generated, and the small bounds on flavour changing neutral currents, which have to be avoided. This is achieved by a sizeable renormalisation enhancement of the techniquark condensate.
Apart from this primary purpose, extended technicolour may also stabilise the vacuum alignment \cite{Peskin:1980gc} and make extra Nambu--Goldstone modes (Only three of the technipions serve as longitudinal modes of the weak gauge bosons.) massive enough to avoid direct detection bounds. The ramifications from extended technicolour are definitely already present before the techniquark condensate forms. 
Some of the effects connected to the stabilisation of the vacuum alignment and the Nambu--Goldstone masses are similar in nature to an explicit mass term akin to the electroweakly induced quark masses relative to the chiral dynamics of quantum chromodynamics.

Therefore, in phenomenological applications the question after exact conformality is somewhat academic: The dynamics of ``pure" technicolour are perturbed due to the coupling to the electroweak and extended technicolour. If the technipions are only to be removed outside detection bounds these modifications can be moderate or even altogether zero \cite{Dietrich:2009ix}. If, however, some of the technipions are to be heavy enough to serve as dark matter candidates, the modification is substantial. It is possible that these perturbations influence the amount of walking of a theory or even make a theory conformal, which otherwise would be exactly conformal.

As just discussed, in ``massless" technicolour theories, walking is caused by intrinsically non-perturbative dynamics. As was already pointed out in \cite{Dietrich:2009ns}, there is also the possibility to see walking from more perturbative effects: Consider an asymptotically free gauge theory that for massless flavours would feature an infrared fixed point but where some of the fermion have ``hard" masses in analogy to the electroweak masses as seen by quantum chromodynamics. Then the infrared fixed point is never reached because the heavy fermions will freeze out for low energy scales, but, depending on the initial conditions, can be approached very closely. We are going to study this circumstance closely in what follows. 

The plateau in the evolution of the coupling is the eponym for walking theories. The feature of importance for the construction of viable technicolour models is the renormalisation of the mass operator of the fermions, $\int^{\mu_2}_{\mu_1}\frac{d\mu}{\mu}\gamma(\mu)$, where $\gamma$ stands for the anomalous dimension of said operator and $\mu$ for the energy scale. Therefore, we will concentrate on this quantity. 

The paper is organised as follows:
Sect.~\ref{BETA} presents our mass-dependent all-order $\beta$-function. 
Sect.~\ref{CONF} analyses the interplay between quasiconformality and mass, where
Sect.~\ref{CONF2} is concerned with the determination quasiconformal window. 
In Sect,~\ref{UNIV} identifies and analyses universal behaviour in the vicinity of a would-be fixed point. 
Sect.~\ref{USE} put our findings to use in the construction of walking technicolour models.
Sect.~\ref{SUMM} summarises the results.


\section{$\beta$-functions\label{BETA}}

The $\beta$-function is defined as the change of the gauge coupling $g$ of a field theory with the energy scale $\mu$ due to renormalisation. In mass-independent subtraction schemes it is scheme independent up to two loops,
\begin{equation}
\beta (g)=-\frac{\beta_0}{(4\pi)^2}g^3-\frac{\beta_1}{(4\pi)^4}g^5-\dots ,
\label{pb}
\end{equation}

\begin{equation}
\beta_0=\frac{11}{3}C_2(G)-\frac{4}{3}T(R)N_f ,
\end{equation}

\begin{equation}
\beta_1=\frac{34}{3}C_2(G)^2-\frac{20}{3}C_2(G)T(R)N_f-4C_2(R)T(R)N_f .
\end{equation}
In the mass-dependent background field momentum subtraction scheme, the first two  coefficients are given by \cite{Jegerlehner:1998zg},
\begin{eqnarray}
\beta_0\mapsto\bar{\beta}_0&=&\frac{11}{3}C_2(G)-\frac{4}{3}T(R)\sum_{j=1}^{N_f}b_0(x_j) ,\label{barbeta0}\\
\beta_1\mapsto\bar{\beta}_1&=&\frac{34}{3}C_2(G)^2-T(R)\sum_{j=1}^{N_f}b_1(x_j) ,
\label{barbeta1}
\end{eqnarray}
where,
$
x_j=-\mu^2/(4m_j^2)
$
and the mass $m_j$ of the fermion flavour $j$  of the fermion flavour $j$. Additionally,
\begin{equation}
b_0(x)=1+3[1-G(x)]/(2x) ,
\label{b0}
\end{equation}
which is gauge invariant \cite{Brodsky:1999fr}. Here
$
G(x)=(2y\ln y)/(y^2-1)
$,
$
y=(\sqrt{1-1/x}-1)/(\sqrt{1-1/x}+1)
$,
\begin{widetext}
\begin{eqnarray}
b_1(x)
&=&
\frac{16(1-x^2)C_2(R)+(1+8x^2)C_2(G)}{6x^2(1-x)}\sigma(x)
-
\frac{2}{3x^2}(C_2(G)-2C_2(R))I(x)
+
\frac{2}{3x}\tilde I_3^{(4)}(x)C_2(G)
+\nn&&+
[(1+3x-10x^2+12x^3)C_2(G)-3(3-3x-4x^2+8x^3)C_2(R)]\frac{4}{3x}G(x)^2
-\nn&&-
[(147-4x-100x^2+8x^3)C_2(G)+168(1-x)C_2(R)+6(9+4x)\ln(-4x)C_2(G)]\frac{1}{9x}G(x)
+\nn&&+
[(99+62x)C_2(G)+12(11+3x)C_2(R)+2(27+24x-2x^2)\ln(-4x)C_2(G)]\frac{1}{9x} ,
\end{eqnarray}
\begin{equation}
\sigma(x)
=
\left\{2\mathrm{Li}_2(-y)+\mathrm{Li}_2(y)+[\ln(1-y)+2\ln(1+y)-(3/4)\ln y]\ln y\right\}
(1-y^2)/y 
\end{equation}
\begin{equation}
I(x)
=
6[\zeta_3+4\mathrm{Li}_3(-y)+2\mathrm{Li}_3(y)]-8[2\mathrm{Li}_2(-y)+\mathrm{Li}_2(y)]\ln y-2[2\ln(1+y)+\ln(1-y)]\ln^2y ,
\end{equation}
\begin{equation}
\tilde I^{(4)}_3(x)
=
6\zeta_3-6\mathrm{Li}_3(y)+6\mathrm{Li}_2(y)\ln y+2\ln(1-y)\ln^2y ,
\end{equation}
\end{widetext}
$\zeta_3=\zeta(3)=1.2020569\dots$, and
$\mathrm{Li}_n(z)$
is the polylogarithm.

For $N_f$ mass degenerate flavours, here, for simplicity, all transforming under the same representation of the gauge group, the modifications of $\bar{\beta}_0$ and $\bar{\beta}_1$ relative to $\beta_0$ and $\beta_1$ can be gathered in ``numbers of active flavours",
\begin{eqnarray}
N_{f,0}&=&N_fb_0(x) ,\\
N_{f,1}&=&N_fb_1(x)/[(20/3)C_2(G)+4C_2(R)] .
\end{eqnarray}
In both cases $N_{f,i}\rightarrow N_f$ for $m\rightarrow 0$ and $N_{f,i}\rightarrow 0$ for $m\rightarrow\infty$. Hence, the decoupling theorem \cite{Appelquist:1974tg} is satisfied.  Further, while $N_{f,0}$ interpolates monotonously between the two limiting cases $N_f$ and zero, which would be expected from a number of active flavours. $N_{f,1}$ does not and is, in general, not even positive for all values of $x$. Hence, the interpretation as active number of flavours is more appropriate for $N_{f,0}$ than it is for $N_{f,1}$. Additionally, these modified numbers of flavours can be given a gauge invariant meaning \cite{Brodsky:1999fr}, which makes the question why there should be different numbers of active flavours in each term of the $\beta$-function even more acute.

Exploiting two-loop universality in the massless case (An expansion to two loops of the following expression reproduces the universal two-loop coefficients.), a massless all-order $\beta$-function was conjectured in \cite{Ryttov:2007cx},
\begin{equation}
\beta(g)
=
-
\frac{g^3}{(4\pi)^2}
\frac{\beta_0-\frac{2}{3}T(R)N_f\gamma(g^2)}{1-\frac{g^2}{8\pi^2}C_2(G)(1+2\frac{\beta^\prime_0}{\beta_0})} .
\label{fsaob}
\end{equation}
Here
$
\beta^\prime_0=C_2(G)-T(R)N_f 
$
and
$
\gamma=-d\ln m/d\ln\mu
$
stands for the anomalous dimension of the fermion mass operator. To universal massless one-loop order
$
\gamma(g^2)=(3/2)C_2(R)g^2/(4\pi^2)+O(g^4).
$
Furthermore, (\ref{fsaob}) reproduces limiting cases in which the $\beta$-function is known exactly like, for example, super Yang--Mills theory \cite{Novikov:1983uc} or planar equivalence in a large-$N_c$ limit \cite{Armoni:2003gp}.

In \cite{Dietrich:2009ns}, our mass-dependent all-order $\beta$-function was derived based on the following requirements: When expanded to two-loop order the $\beta$-function coefficients in the background field momentum subtraction scheme are to be reproduced. \{Alternatively, we could have used the physical charge approach of Ref.~\cite{Brodsky:1999fr} as target. The outcome, however, is qualitatively and quantitatively close to that in the background field momentum subtraction scheme (See especially Fig.~2 in \cite{Brodsky:1999fr}.) and in the latter the expressions can be handled analytically.\} Further, in the massless limit the mass-dependent all-order $\beta$-function is to coincide with the mass-independent all-order $\beta$-function. This ensures also that the exactly known results from supersymmetry are reproduced. Finally, in the ultramassive limit it is to coincide with the pure Yang--Mills version of the mass-independent all-order $\beta$-function. This implies that all terms involving the Casimir $C_2(R)$ have to be absorbed in the term involving the anomalous dimension. This resulted in the following mass-dependent $\beta$-function,
\begin{equation}
\bar{\beta}(g)
=
-
\frac{g^3}{(4\pi)^2}
\frac{\bar{\beta}_0-\frac{2}{3}T(R)\sum_{j=1}^{N_f}\bar{\gamma}(x_j)}{1-\frac{g^2}{8\pi^2}C_2(G)(1+2\frac{\bar{\beta}^\prime_0}{\bar{\beta}_0})} .
\label{maob}
\end{equation}
Here
\begin{equation}
\bar{\gamma}(x_j)=
{\frac{3}{2}\frac{g^2}{4\pi^2}}\frac{1}{4}b_1(x_j)|_{C_2(G)\rightarrow 0}
\end{equation}
and
\[
\bar{\beta}^\prime_0
=
C_2(G)
+
T(R)\sum_{j=1}^{N_f}\left[
\frac{2}{3}b_0(x_j)
-
{
\frac{1}{4}\frac{b_1(x_j)|_{C_2(R)\rightarrow 0}}{C_2(G)}
}
\right] .
\]
It can be generalised to accommodate flavours that transform under different representations of the gauge group. Then it becomes
\begin{equation}
\bar{\beta}(g)
=
-
\frac{g^3}{(4\pi)^2}
\frac{\bar{\beta}_0-\frac{2}{3}\sum_{j=1}^{N_f}T(R_j)\bar{\gamma}(x_j)}{1-\frac{g^2}{8\pi^2}C_2(G)(1+\frac{\bar{\beta}^\prime_0}{\bar{\beta}_0})} .
\end{equation}
Here
\begin{equation}
\bar{\beta}^\prime_0=C_2(G)+\sum_{j=1}^{N_f}T(R_j)\left[\frac{2}{3}b_0(x_j)-
{
\frac{1}{4}\frac{b_1(x_j)|_{C_2(R)\rightarrow 0}}{C_2(G)}
}\right]
\end{equation}
and $R_j$ is the representation of the flavour $j$. At fixed coupling $\bar{\gamma}(x_j)$ goes to 0 for $x\rightarrow 0$ and to the massless value for $x\rightarrow-\infty$, which is also observed for the mass-dependent anomalous dimension in \cite{Georgi:1976ve}.
\{As was discussed before in \cite{Dietrich:2009ns}, this would still be the case if a factor $1+O(g^2)$ was incorporated in $\bar{\gamma}$ and/or $\bar{\beta}_0^\prime$. It would make appearance at third order and could be absorbed in a change of the renormalisation scheme, which is already true in the massless case. Hence, if one wanted to accommodate a particular three-loop term, one could include such a factor and adjust the denominator accordingly, which would amount to a change of scheme. One particular massless scheme adapted for studies in the framework of holographic duals \cite{Antipin:2009dz} was introduced in \cite{Antipin:2009wr}.\}


\section{Quasiconformality and mass\label{CONF}}

Strictly speaking, already the exactly supersymmetric $\beta$-function constitutes ``only" a relation between the $\beta$-function and the anomalous dimension of the fermion mass operator, and the latter is not known to all orders. Therefore, we do not have a parametrisation of the $\beta$-function of the form $\beta=\beta(g)$ at our disposal. This circumstance is inherited by the supersymmetry-inspired massless and mass-dependent $\beta$-functions.


\subsection{The (quasi)conformal window\label{CONF2}}

Despite our ignorance of the exact form of the anomalous dimension, the all-order $\beta$-functions allow us to find the lower bound of the conformal window or, in the mass-dependent case, the quasiconformal window: 

In Ref.~\cite{Dietrich:2006cm}, the lower bound of the conformal window in the $N_c$-$N_f$ plane was determined by equating the coupling at the Caswell--Banks--Zaks fixed point with the critical coupling \cite{Appelquist:1988yc} for the formation of a chiral condensate in the ladder-rainbow-approximation to the Dyson--Schwinger--equations. 

In Ref.~\cite{Ryttov:2007cx}, Eq.~(\ref{fsaob}) was used to determine the lower bound of the conformal window by setting it equal to zero while putting the value of the anomalous dimension to its critical value for the onset of chiral symmetry breaking. The ladder-rainbow-approximation yields 1 as critical value. The only known theoretically hard upper bound on the anomalous dimension, however, arises from the requirement of unitarity of the field theory and is 2. (This is a consequence of the fact, that in a conformal field theory the dimension $3-\gamma$ of all non-trivial spinless operators including that of the chiral condensate must be larger or equal to unity \cite{Mack:1975je} to avoid negative norm states.) The lower bound for the conformal window must, hence, not lie at a lower number of flavours. (Duality arguments \cite{Sannino:2009qc} also give indications for choosing the critical value for the anomalous dimension.) This method has also been used for gauge groups other than $SU(N)$ and for multiple representations \cite{Sannino:2009aw,Ryttov:2009yw}.
The framework leads to a universal relation for the lower bound of the conformal window \cite{Armoni:2009jn},
$
1=\kappa~2N_f~T(R)/C_2(G) .
$
In Ref.~\cite{Armoni:2009jn}, this relation was found in the worldline formalism; the value $\kappa\approx 1/4$ was determined from matching to SQCD. For comparison, from Eq.~(\ref{fsaob}), one finds $\kappa=(2+\gamma)/11$. (A combination of the two results would yield $\gamma\approx 3/4$.) An analogous relation was also found in \cite{Poppitz:2009uq}.

In Ref.~\cite{Dietrich:2009ns}, we studied the influence of threshold effects due to finite---that is, neither formally zero nor infinite---fermion masses based on the mass-dependent all-order $\beta$-function (\ref{maob}) by fixing the anomalous dimension to the two benchmark values from the massless study.
In a theory, where all fermion masses are nonzero, they freeze out for scales far enough below this mass. As a consequence, we are then effectively left with a pure Yang--Mills theory, where the antiscreening from the gluons is uncompensated.
Therefore, what is determined by applying the above-described formalism is the phenomenologically decisive minimal number of flavours above which the coupling develops a plateau, that is, walks. Thus, we talk of a quasiconformal window, which in the massless limit coincides with the conformal window.
For nonzero masses,
the lower bound of the quasiconformal window is shifted to a larger number of flavours.
In the strictly massless case, the walking theories are found slightly below the lower edge of the conformal window; above the edge, the theory evolves into the fixed point.
In the massive case, slightly below the lower bound of the quasiconformal window, there will also be at least some walking. 
We expect that the amount of walking---the range of scales of quasiconformal behaviour---is determined by an interplay of the freezing out of the flavours due to the explicit mass and the onset of chiral symmetry breaking. (Independent of walking, a similar interplay between quark mass effects and chiral symmetry breaking exists in quantum chromodynamics for the strange quark.) 
For a number of flavours too far below this bound the theory never comes close to the would-be fixed point and does not exhibit any walking. For a number of flavours above the lower bound of the quasiconformal window an asymptotically free theory approaches the fixed point very closely and stays in its vicinity until the flavours start freezing out gradually. Once the flavours are decoupled sufficiently, the coupling starts running again. (We are going to study this decoupling process below.) As a consequence, the position of the low-energy end of the plateau is not set by the initial conditions for the renormalisation group evolution of the gauge coupling alone, but also by the value of the fermion masses. 
In Ref.~\cite{Dietrich:2009ns} we have determined the quasi-conformal windows for $SU(N_c)$, $Sp(2N_c)$, and $SO(N_c)$ gauge groups.
The critical number of flavours obtained by setting the mass-dependent $\beta$-function equal to zero at a fixed critical value $\gamma_c$ of the anomalous dimension is given by,
\begin{equation}
N_f=\frac{11}{2}\frac{C_2(G)}{T(R)}[\gamma_c+2b_0(x)]^{-1} .
\label{criterion}
\end{equation}
The modification due to the mass of the fermion is, hence, universal in the sense that it neither depends on the gauge group nor the representation. The latter is encoded in the fraction 
$C_2(G)/T(R)$, while the mass effect is contained in $b_0(x)$. 


\subsection{Universal evolution in the quasiconformal window\label{UNIV}}

\begin{figure*}[t]
\hfill\subfigure{\resizebox{!}{5.8cm}{\includegraphics{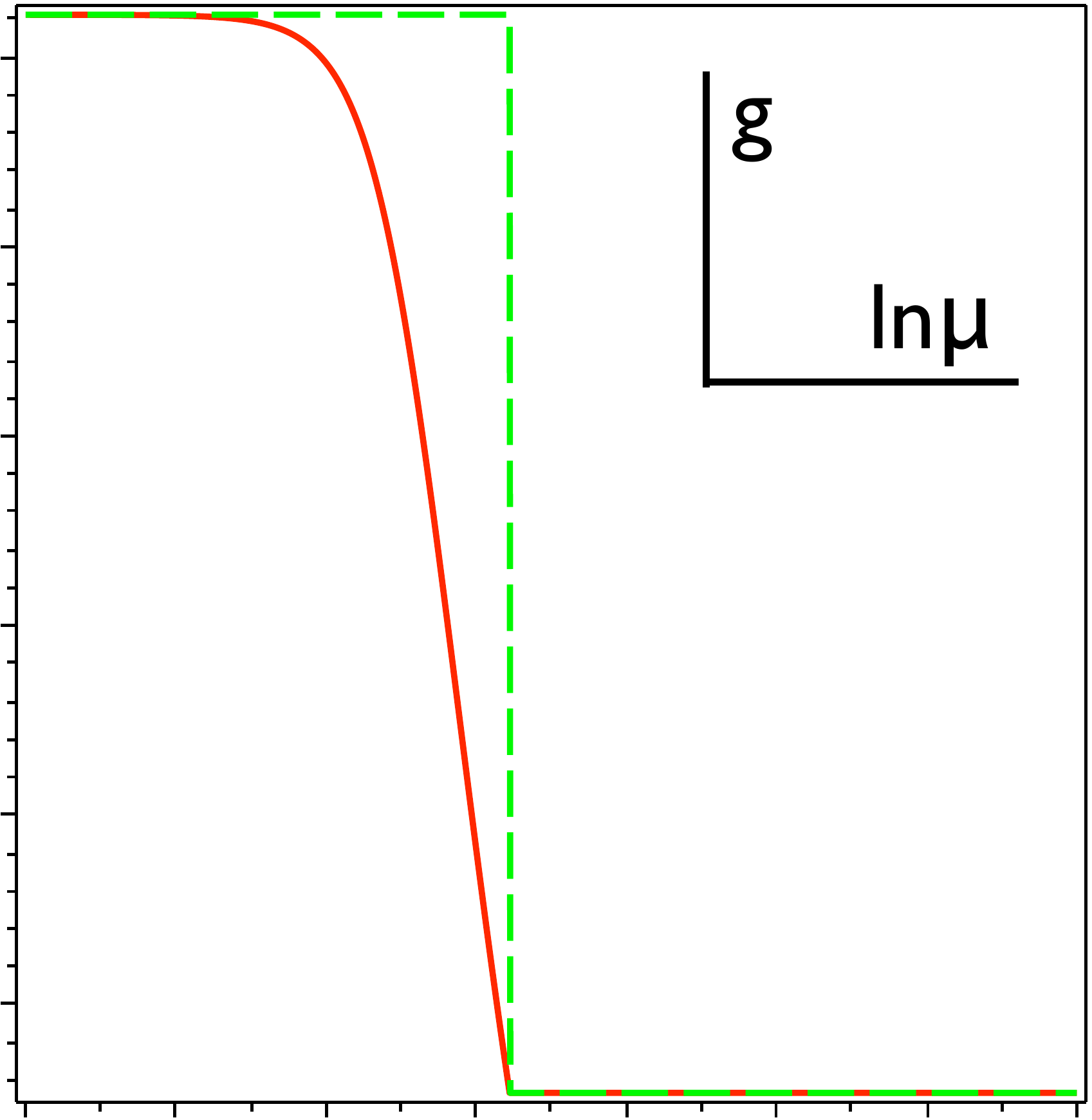}}}\hfill\subfigure{\resizebox{!}{5.8cm}{\includegraphics{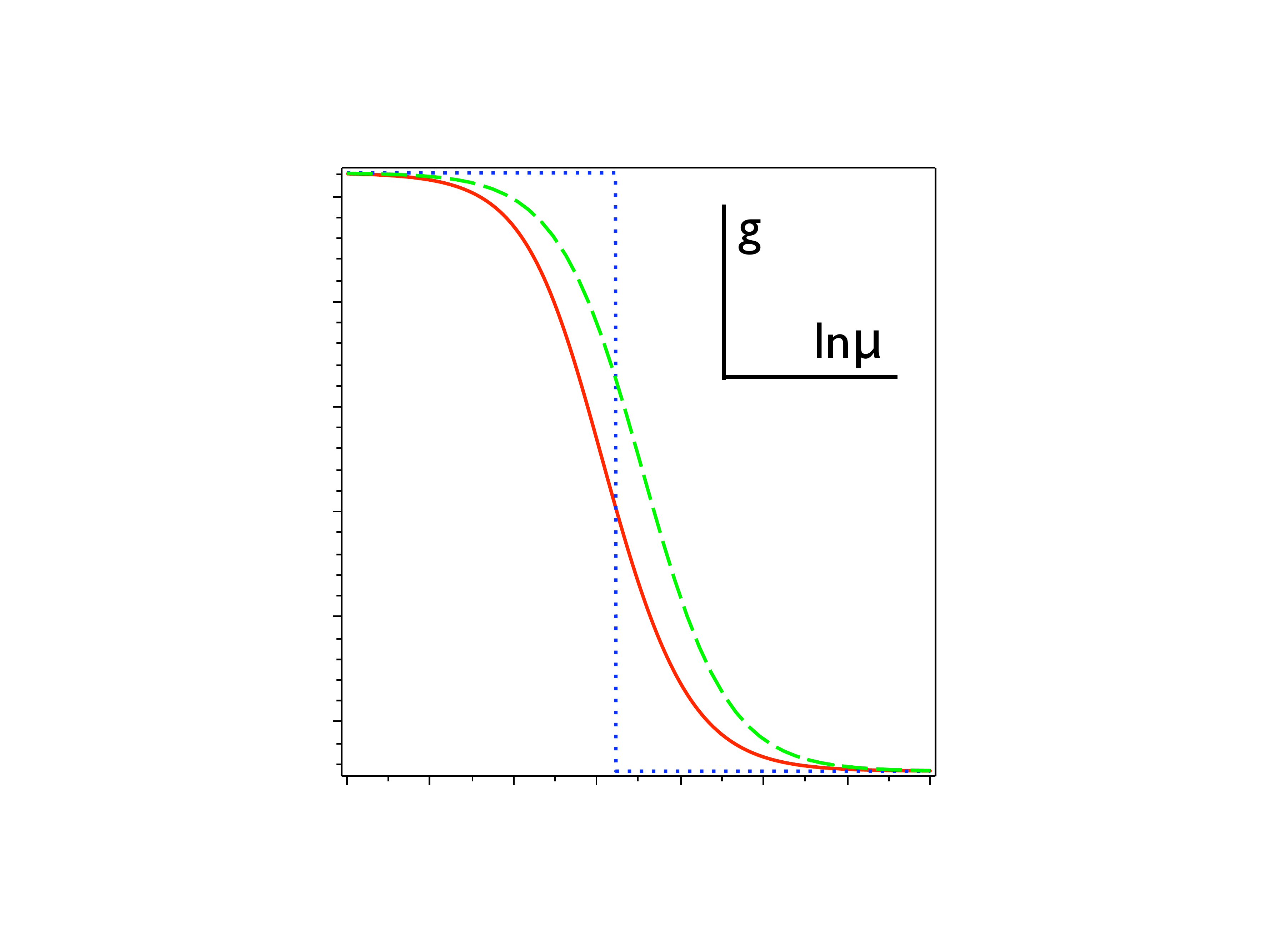}}}\hfill~
\caption{
Perturbative two-loop study of $g=g(\ln\mu)$ for the switching off of one flavour. 
Relaxation of the coupling $g$ to the new fixed point value of the coupling (solid, red).
Fixed point value of the coupling for the active number of flavours at a given value of $\ln\mu$, $g[N_f(\ln\mu)]$ (dashed green).
Left panel: sudden switching.
Right panel: smooth switching; the blue dotted line serves to indicate twice the mass of the flavour.
} 
\label{freeze}
\end{figure*}

Here we extract further pieces of information on the evolution of quasiconformal gauge field theories from the mass-dependent all-order $\beta$-function (\ref{maob}). As we do not know the exact expression for the anomalous dimension of the fermion mass operator, we cannot describe the renormalisation group evolution for just any generic setup. There is, however, universal behaviour in the vicinity of the would-be fixed point: Consider a theory inside the quasiconformal window, where the evolution has (almost) reached what would be the infrared fixed point in the complete absence of mass. In that case, the $\beta$-function is almost zero, as is its numerator. (At the same time, the coupling $g$ is nonzero and thus is the anomalous dimension. Hence, while the coupling is almost stationary---due to the small value of $\beta$---the mass still grows when we decrease the energy scale.)
When we reduce the energy scale, massive flavours will (continue to) freeze out, which leads to an increase of $\bar\beta_0$.
As a consequence, the $\beta$-function becomes slightly more negative, which, in turn, makes the coupling $g$ grow slowly with decreasing energy scale. For an anomalous dimension that is a growing function of the coupling $g$ the $\beta$-function becomes, thus, again less negative. Taking stock, we have two counteracting effects, one---the freezing out of flavours---which drives $\beta$ to negative values, and a second---the slow increase of the anomalous dimension---which restores the $\beta$-function to zero. 

From the definition
\begin{equation}
\delta=\bar{\beta}_0-\frac{2}{3}T(R)\sum_{j=1}^{N_f}\bar{\gamma}(x_j) 
\label{delta}
\end{equation}
we get
\begin{equation}
\bar{\beta}(g)
=
-
\frac{g^3}{(4\pi)^2}
\frac{\delta}{1-\frac{g^2}{8\pi^2}C_2(G)(1+2\frac{\bar{\beta}^\prime_0}{\bar{\beta}_0})} 
\label{betadelta}
\end{equation}
and, considering for the moment $N_f$ mass degenerate flavours,
\begin{equation}
\bar{\gamma}=\frac{11}{2}\frac{C_2(G)}{T(R)N_f}-2b_0-\frac{3}{2}\frac{\delta}{T(R)N_f} .
\label{gammadelta}
\end{equation}
For the freeze out of a flavour as described by $b_0$ the switching zone spans five orders of magnitude. (This also means, threshold effects are felt for energy scales, which are more than a hundred times bigger than the mass of the fermions.)
We expect that the growth of the anomalous dimension and, hence, the restoration of the $\beta$-function to almost zero, is able to follow nearly ``adiabatically" the very gradual freezing out of flavours over four to five orders of magnitude as described by $b_0$. This can at least be seen from the two-loop study depicted in Fig.~\ref{freeze}: When switching off a flavour suddenly (left panel) the coupling deviates strongly from its fixed point values. If it is switched off gradually, as described by $b_0$ (right panel), the coupling is almost able to follow its fixed point value. As a consequence, $\delta$ will be small and positive in the second case. 
Hence, in the following, we will at first neglect $\delta$ and will determine the corrections arising from a finite $\delta$ afterwards. 
Thus, omitting $\delta$ for the moment, we are left with the differential equation
\begin{equation}
-\frac{d\ln m}{d\ln\mu}
=
\frac{11}{2}\frac{C_2(G)}{T(R)N_f}-2b_0\left(\frac{m}{\mu}\right) .
\label{gammawodelta}
\end{equation}
With $b_0$ from Eq.~(\ref{b0}) an analytic integration is not possible. Therefore, we will use the approximation \cite{Brodsky:1999fr} 
\begin{equation}
b_0\approx(1+5m^2/\mu^2)^{-1},
\label{approx}
\end{equation}
which deviates by at most $\approx1\%$ from Eq.~(\ref{b0}) over the entire range of scales, and find
\begin{equation}
(c-1)\ln\frac{\mu}{\mu_0}
=
\ln\frac{\frac{m_0}{\mu_0}}{\frac{m}{\mu}}
+\frac{1}{c+1}
\ln\frac{c-1+5(c+1)\frac{m^2}{\mu^2}}{c-1+5(c+1)\frac{m^2_0}{\mu^2_0}} ,
\label{massrenorm}
\end{equation}
where 
$c=\frac{11}{2}\frac{C_2(G)}{T(R)N_f}$.
As the lower bounds of integration we choose the point where Eq.~(\ref{criterion}) is satisfied, that is, where the critical anomalous dimension is reached and chiral condensation sets in, such that 
$\frac{m_0^2}{\mu^2_0}=\frac{1}{5}(\frac{2}{c-\gamma_c}-1)$.
 
\begin{figure*}[t]
\hfill\subfigure{\resizebox{!}{5.8cm}{\includegraphics{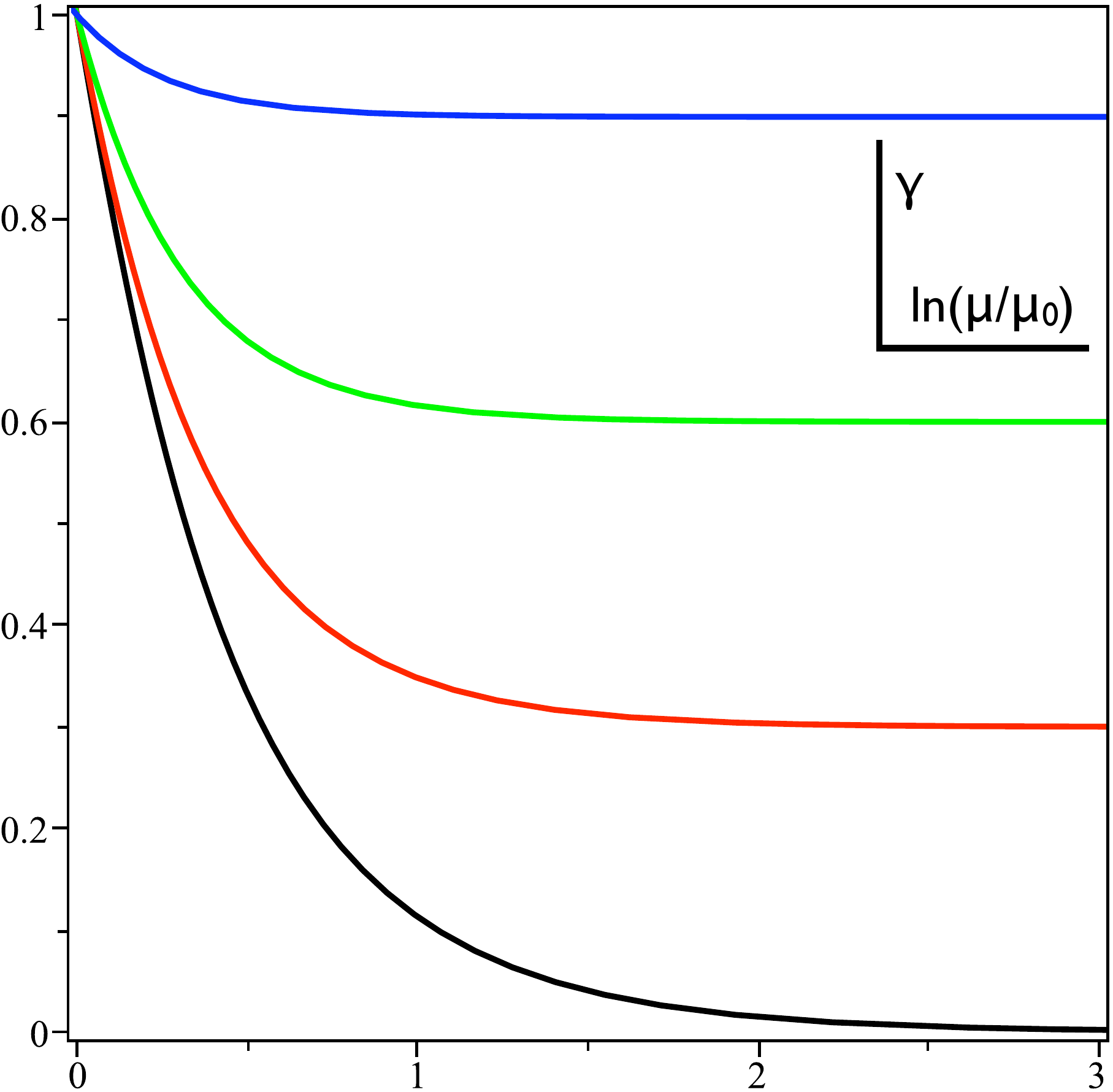}}}\hfill\subfigure{\resizebox{!}{5.8cm}{\includegraphics{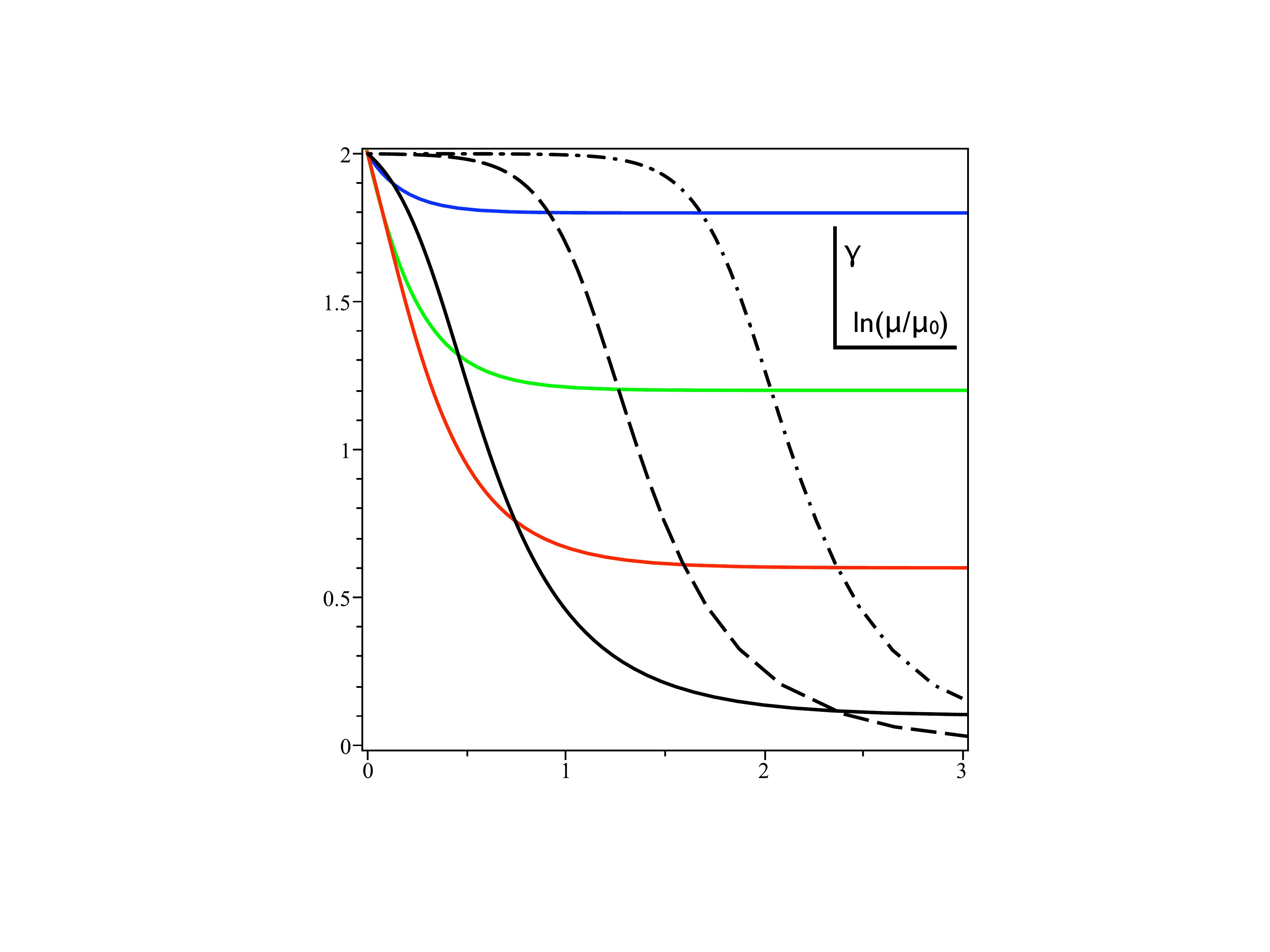}}}\hfill~
\caption{Anomalous dimension as a function of the energy scale: Left panel, $\gamma_c=1$; from top to bottom: $c=2.9$ (blue), $c=2.6$ (green), $c=2.3$ (red), $c=2.0$ (black). Right panel, $\gamma_c=2$; from top to bottom, solid: $c=3.8$ (blue), $c=3.2$ (green), $c=2.6$ (red), $c=2.1$ (black); Not solid, $c=2.001$ (dashed), $c=2.00001$ (dash-dotted). 
} 
\label{gamma_mu}
\end{figure*}

Taking stock, Eq.~(\ref{gammawodelta}) gives the anomalous dimension as a function of the ratio $m/\mu$ and Eq.~(\ref{massrenorm}) the scale $\mu$ as a function of the ratio $m/\mu$. Hence, we are now in the postion to plot the anomalous dimension as a function of the scale $\mu$. (See Fig.~\ref{gamma_mu}) For this we choose again the two customary benchmark values, where we set the critical anomalous dimension equal to 1 (left panel) or to 2 (right panel). We show the result for different values of $c\ge2$, which is required for asymptotic freedom. On the left-hand side for all values of $c$ and on the right-hand side for values of the parameter $c$ far enough above 2, one sees the anomalous dimension comes from a plateau the position $(c-2)$ of which is determined by the condition that the numerator of the $\beta$-function vanish for all flavours active. Once the flavours start freezing out, the anomalous dimension starts increasing until the critical value for chiral condensation is reached. In the case where this critical value is 2, we see also a different behaviour for values of $c$ close to 2: A second plateau exists at the largest values of the anomalous dimension at the end of which the critical value is only just reached. (We will see more examples for this behaviour below.) That such a kind of behaviour can arise is linked to the fact that here we have to consider the limit $m/\mu\rightarrow\infty$ at fixed anomalous dimension and not fixed coupling. While the latter limit leads to the pure Yang--Mills result, the former leads to the result for $N_f/2$ flavours at $b_0=0$ in Eq.~(\ref{gammawodelta}). 

\begin{figure*}[t]
\hfill\subfigure{\resizebox{!}{5.8cm}{\includegraphics{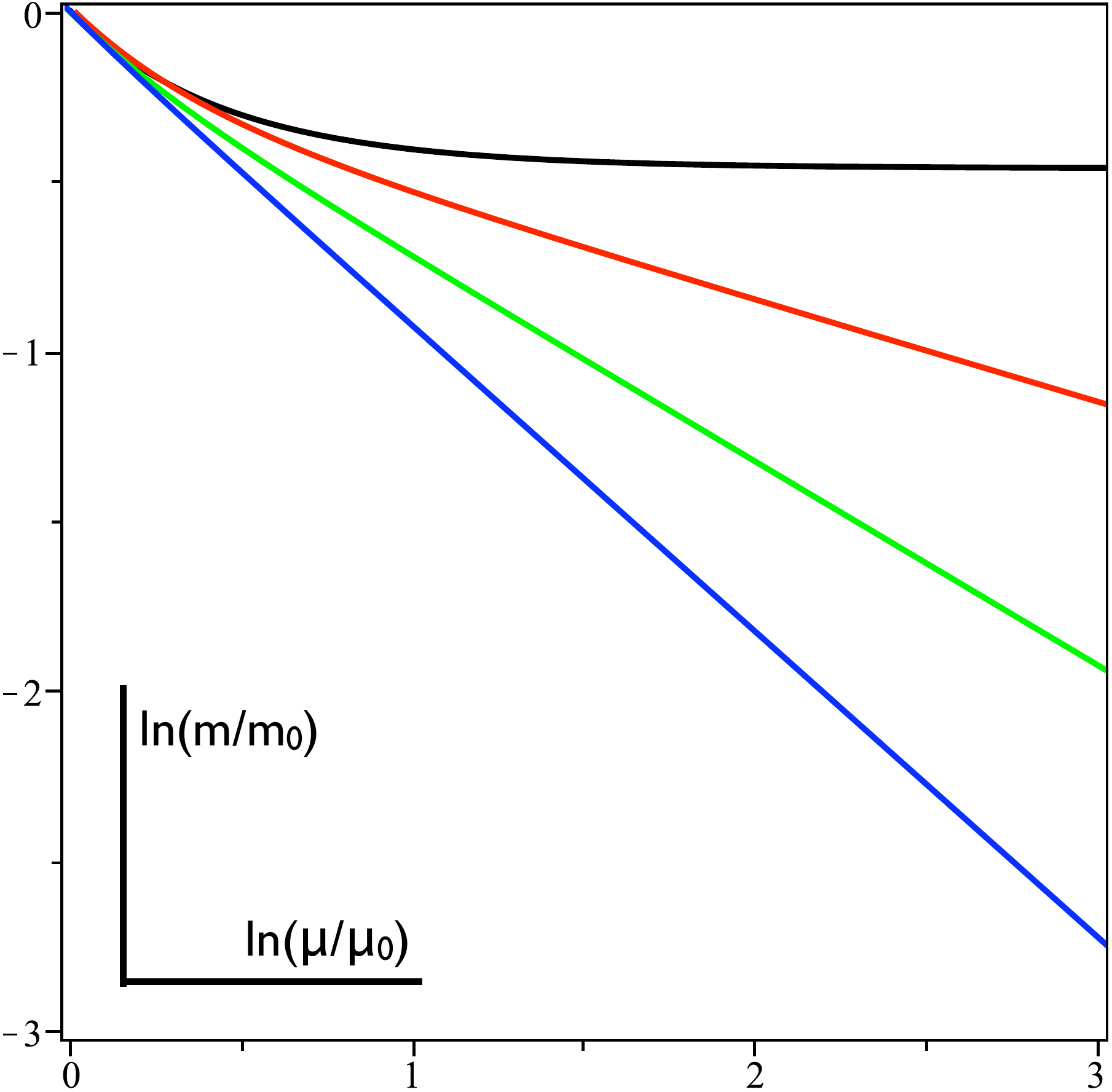}}}\hfill\subfigure{\resizebox{!}{5.8cm}{\includegraphics{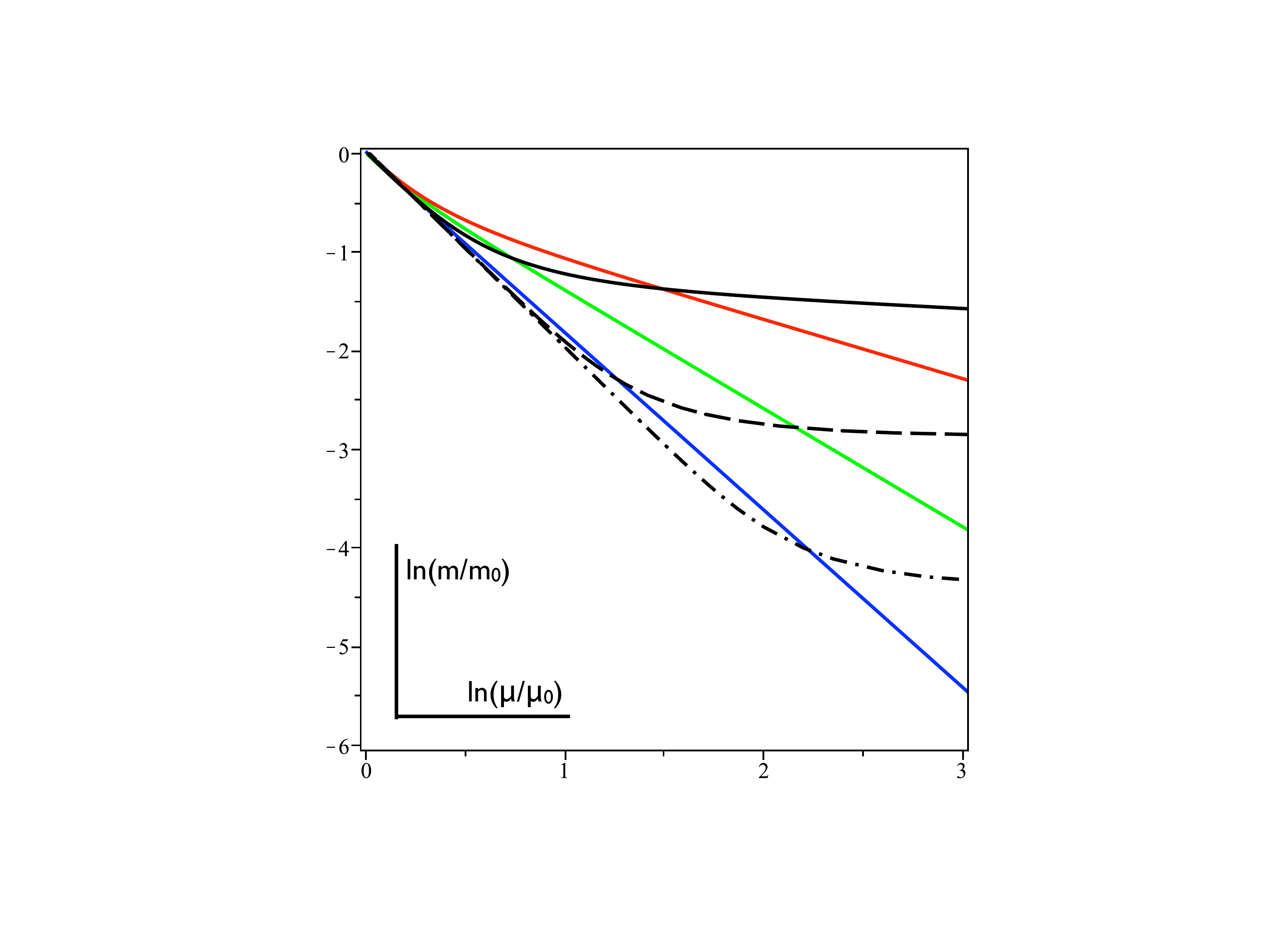}}}\hfill~
\caption{Renormalisation of the mass as a function of the energy scale. Left panel $\gamma_c=1$; from bottom to top: $c=2.9$ (blue), $c=2.6$ (green), $c=2.3$ (red), $c=2.0$ (black). Right panel $\gamma_c=2$; from bottom to top, solid: $c=3.8$ (blue), $c=3.2$ (green), $c=2.6$ (red), $c=2.1$ (black); $c=2.001$ (dashed), $c=2.00001$ (dash-dotted).}
\label{m_mu}
\end{figure*}

Further, we can also plot the logarithm of the mass renormalisation factor, $\ln[m(\mu)/m(\mu_0)]$, as a function of the scale. (See Fig.~\ref{m_mu}.) As would be expected, the cases with pre freeze out plateau, where the plateau is at the largest value of the anomalous dimension, show the biggest renormalisation factor for the mass. Their efficiency is, however, easily rivalled by those cases with a post freeze out plateau, where the anomalous dimension lingers very close to its critical value.

In order to determine the corrections from a finite $\delta$, we invert Eq.~(\ref{betadelta}) and get
\begin{eqnarray}
\delta
&=&
-\frac{(4\pi)^2}{g^3}\left[1-\frac{g^2}{8\pi^2}C_2(G)\left(1+2\frac{\bar{\beta}^\prime_0}{\bar{\beta}_0}\right)\right]\frac{dg}{d\ln\mu}
=\nonumber\\&=&
8\pi^2\frac{dg^{-2}}{d\ln\mu}+2\frac{d\ln g}{d\ln\mu}C_2(G)\left(1+2\frac{\bar{\beta}^\prime_0}{\bar{\beta}_0}\right)
\end{eqnarray}
Integrating and making use of the mean value theorem yields
\begin{equation}
\int_{\mu_1}^{\mu_2} \frac{d\mu}{\mu}  \delta
=
\left[\frac{2\pi^2}{g^2}+\frac{1}{2}\ln(g)C_2(G)\left(1+2\left\langle\frac{\bar{\beta}^\prime_0}{\bar{\beta}_0}\right\rangle\right)\right]_{g(\mu_1)}^{g(\mu_2)} ,
\end{equation}
where $\langle\bar{\beta}^\prime_0/\bar{\beta}_0\rangle$ stands for $\bar{\beta}^\prime_0/\bar{\beta}_0$ evaluated at an intermediate value of $\mu$. 
If the anomalous dimension changes slowly, the $\beta$-function remains very close to zero, $g(\mu_2)$ stays close to $g(\mu_1)$, and the deviation from our result is small. Where the anomalous dimension changes at a higher rate the logarithmic term leads to an enhancement of the mass renormalisation and the $g^{-2}$ term to a reduction. 
Hence, we can estimate,
\begin{equation}
\int_{\mu_1}^{\mu_2} \frac{d\mu}{\mu} ~ \delta
<
\frac{8\pi^2}{[g(\mu_2)]^2}-\frac{8\pi^2}{[g(\mu_1)]^2}
<
\frac{8\pi^2}{[g(\mu_2)]^2} .
\end{equation}
If we accept that for a given value of the coupling the full expression for the anomalous dimension is reduced relative to the one-loop result by the higher order terms, we get 
\begin{equation}
\int_{\mu_1}^{\mu_2} \frac{d\mu}{\mu} ~ \delta
<
\frac{8\pi^2}{[g(\mu_2)]^2}
=
3\frac{C_2(R)b_0}{\gamma_{1\bigcirc}(\mu_2)}
<
3\frac{C_2(R)b_0}{\gamma(\mu_2)} .
\label{30}
\end{equation}
According to Eq.~(\ref{gammadelta}) the number that has to be compared to the renormalisation factor of the mass is given by,
\begin{eqnarray}
\frac{3}{2}\int_{\mu_1}^{\mu_2} \frac{d\mu}{\mu}\frac{\delta}{T(R)N_f}
<
\frac{9}{2}\frac{C_2(R)b_0}{T(R)N_f\gamma(\mu_2)}
\ll\nonumber\\\ll
\int_{\mu_1}^{\mu_2} \frac{d\mu}{\mu} \bar{\gamma}
<
\bar{\gamma}(\mu_2)\ln\frac{\mu_2}{\mu_1} .
\end{eqnarray}
We are interested in theories with small $N_c$ and $N_f$. This makes all quantities in the previous inequality $O(1)$ apart from the factor $b_0$. For large masses $m\gg\mu$ it becomes $b_0\ll 1$. Hence, the previous very conservative estimate suffices to show that our computation is accurate not only in the post freeze out plateau but also in the part of the slope where $b_0\ll1$. For assessing the cases with pre freeze out plateaus the last estimate in Eq.~(\ref{30}) is to lavish. For a pre freeze out plateau, if the value of $\int(1-b_0)d\ln\mu \ll 1$ one is on the safe side. All the way to the onset of chiral condensation this criterion is only satisfied for large values of $c$. (See, for example the blue lines in Fig.~\ref{gamma_mu}).

But what is the effect of a finite $\delta$? In regions of changing $b_0$ the effective value of $c$ is reduced. The curves in the right panel of Fig.~\ref{gamma_mu} get steeper as $c$ becomes smaller. Hence, addressing a finite $\delta$ would generically steepen the slopes there more, while leaving the plateaus unchanged. The same holds for the red, green, and blue curves in the right panel. For the black curves, however, the effective reduction of $c$ does not lead to any steeper slopes, but the post freeze out plateau is stabilised by the reduction of $c$. 

\subsubsection{Partially massive}

Now let us generalise to the case with $N_f$ massive and $n_f$ massless flavours. Eq.~(\ref{delta}) becomes
\begin{equation}
\delta
=
\frac{11}{3}C_2(G)
-
\frac{2}{3}T(R)[N_f(2b_0+\bar{\gamma})+n_f(2+\hat{\gamma})] ,
\end{equation}
where $\hat\gamma$ is the anomalous dimension for the massless fermions.
For a given value of the coupling $g$, the anomalous dimension of the massive fermions is reduced relative to that of the massless fermions and at fixed coupling goes to zero in the limit of infinite masses. Defining the ratio between the two as
$
\hat{\gamma}=\bar{\gamma}/\hat{b}_0
$
leads to the equivalent of Eq.~(\ref{gammawodelta}),
\begin{equation}
\bar{\gamma}
=
\frac{11}{2}\frac{C_2(G)}{T(R)(N_f+n_f/\hat{b}_0)}
-
2\frac{N_fb_0+n_f}{N_f+n_f/\hat{b}_0} .
\label{a}
\end{equation}
For $b_0$ and $\hat{b}_0$ close to unity, $b_0=1-\epsilon$ and $\hat{b}_0=1-\hat{\epsilon}$, that is, for small masses, we find
\begin{eqnarray}
\bar{\gamma}
&\approx&
\frac{11}{2}\frac{C_2(G)}{T(R)(N_f+n_f)}\left(1-\frac{n_f\hat{\epsilon}}{N_f+n_f}\right)
-\\&&-
2\left(1-\frac{N_f\epsilon+n_f\hat{\epsilon}}{N_f+n_f}\right) .
\label{b}
\end{eqnarray}
For $b_0$ and $\hat{b}_0$ close to zero, that is, for large masses,
\begin{equation}
\bar{\gamma}
\approx
\left[\frac{11}{2}\frac{C_2(G)}{T(R)n_f}
-
2\right]\hat{b}_0 .
\end{equation}
Hence, $b_0$ and $\hat{b}_0$ can only be small if the anomalous dimension for the massive flavours is small. In fact, the previous relation gives consistently the known fixed point value of the anomalous dimension for $n_f$ massless flavours,
\begin{equation}
\hat{\gamma}
\approx
\frac{11}{2}\frac{C_2(G)}{T(R)n_f}
-
2 .
\end{equation}
Expanding to the next order we get,
\begin{equation}
\hat{\gamma}
\approx
\left(\frac{11}{2}\frac{C_2(G)}{T(R)n_f}-2\right)\left(1-\frac{N_f}{n_f}\hat{b}_0\right)-2\frac{N_f}{n_f}b_0 .
\label{c}
\end{equation}
We notice that in Eq.~(\ref{a}), as is visible more directly in Eqs.~(\ref{b}) and (\ref{c}), $b_0$ and $\hat{b}_0$ always act in the same direction; everywhere appear weighted
sums of the two reduction factors and never differences. Therefore, the outcome is less sensitive to the exact form of $b_0$ relative to $\hat{b}_0$ as the general requirements are satisfied. Therefore, we use henceforth $\hat{b}_0=b_0$. Thus, 
\begin{equation}
\bar{\gamma}
=
\left[\frac{11}{2}\frac{C_2(G)}{T(R)(N_fb_0+n_f)}
-
2\right]b_0 ,
\label{39}
\end{equation}
or
\begin{equation}
\hat{\gamma}
=
\frac{11}{2}\frac{C_2(G)}{T(R)(N_fb_0+n_f)}
-
2 .
\label{criterion2}
\end{equation}
Hence, we can write the equivalent of Eq.~(\ref{gammawodelta}) with $b_0$ from Eq.~(\ref{approx}) as
\begin{equation}
d\ln\mu
=
\frac
{-(1+5\frac{m^2}{\mu^2})[N_f+n_f(1+5\frac{m^2}{\mu^2})]d\frac{m}{\mu}/\frac{m}{\mu}}
{
(1+5\frac{m^2}{\mu^2})C
-
(1-5\frac{m^2}{\mu^2})[N_f+n_f(1+5\frac{m^2}{\mu^2})]
} ,
\end{equation}
where $C=\frac{11}{2}\frac{C_2(G)}{T(R)}$. The integral can be carried out analytically but leads to a rather lengthy expression, which we choose not to display here.
%
\begin{figure*}[t]
\hfill\subfigure{\resizebox{!}{5.8cm}{\includegraphics{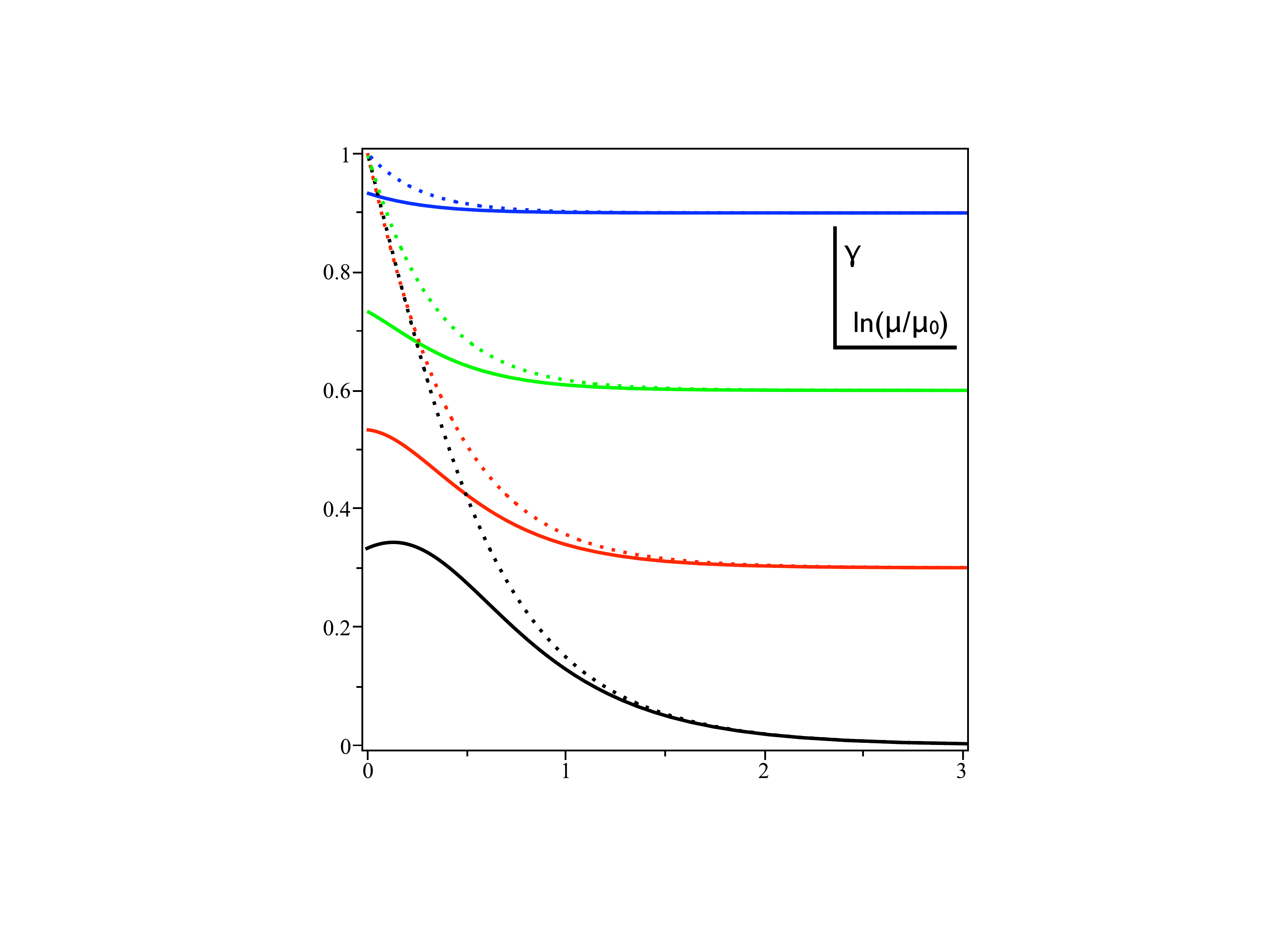}}}\hfill\subfigure{\resizebox{!}{5.8cm}{\includegraphics{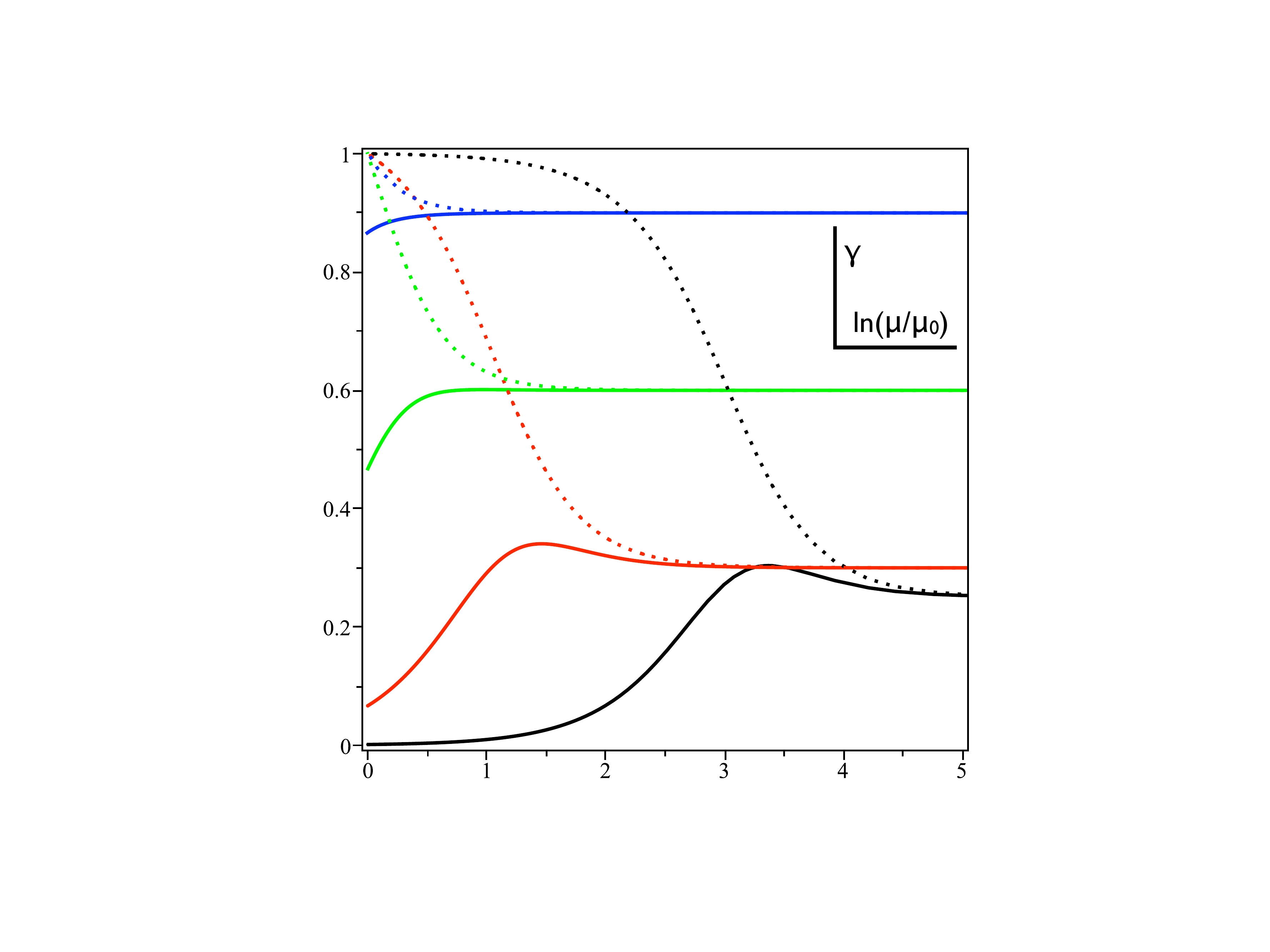}}\hfill\subfigure{\resizebox{!}{5.8cm}{\includegraphics{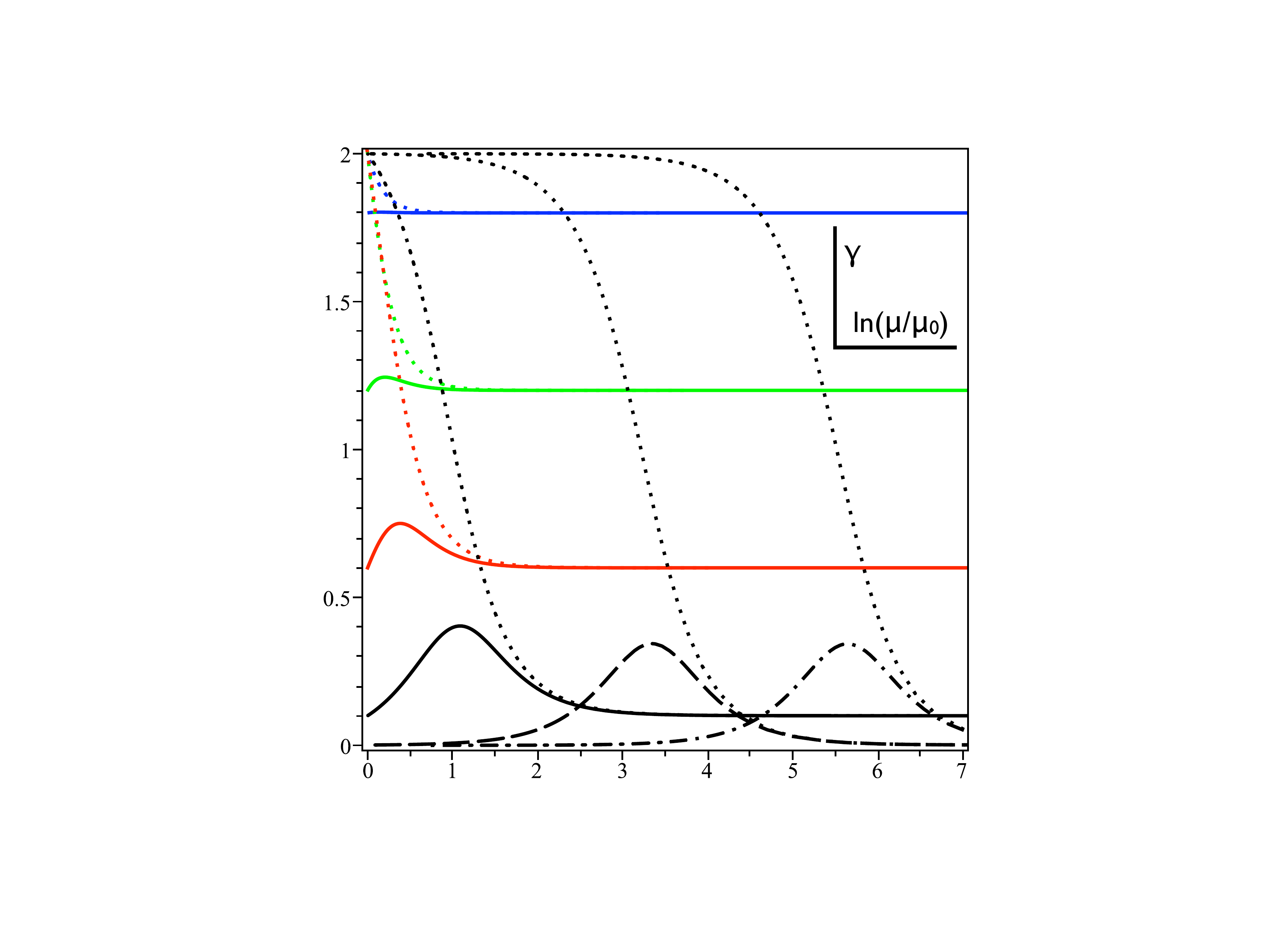}}}}\hfill~
\caption{Anomalous dimension of the massless (dotted) and massive (other) fermions as function of the energy scale. The plateaus at high scales level out at $\gamma=c-2$. {\it Left panel}: $\gamma_c=1$, $N_f=n_f$; from top to bottom, $c=C/(N_f+n_f)=2.9$ (blue), $c=2.6$ (green), $c=2.3$ (red), $c=2.0$ (black). {\it Middle panel}: $\gamma_c=1$, $3N_f=n_f$; from top to bottom, $c=2.9$ (blue), $c=2.6$ (green), $c=2.3$ (red), $c=2.251$ (black). {\it Right panel}: $\gamma_c=2$, $N_f=n_f$; from top to bottom, $c=3.8$ (solid blue), $c=3.2$ (solid green), $c=2.6$ (solid red), $c=2.1$ (solid black); $c=2.001$ (black dashed), $c=2.00001$ (black dash-dotted). 
} 
\label{gamma_mu_p}
\end{figure*}
%
As lower boundary of integration we choose the point where Eq.~(\ref{criterion2}) equals the critical value of the anomalous dimension for which chiral condensation sets in, such that 
\begin{equation}
5\frac{m_0^2}{\mu_0^2}
=
\left[\frac{11}{2}\frac{C_2(G)}{T(R)N_f}\frac{1}{\gamma_c+2}-\frac{n_f}{N_f}\right]^{-1}-1 .
\end{equation}

In the massless sector we get from $\hat{\gamma}=-d\ln\lambda/d\ln\mu$ using the chain rule,
\begin{equation}
\frac{d\ln\lambda}{d\ln\frac{m}{\mu}}=\frac{\hat{\gamma}}{1+\bar{\gamma}} .
\end{equation}
Making use of Eqs.~(\ref{39}) and (\ref{criterion2}), we obtain
\begin{equation}
\frac{d\ln\lambda}{d\ln\frac{m}{\mu}}
=
\frac{
[b_0^{-1}
(C-2n_f)-2N_f]b_0^{-1}
}{
n_fb_0^{-2}
+(N_f-2n_f+C)b_0^{-1}
-2N_f
} ,
\end{equation}
which can be integrated analytically after making use of Eq.~(\ref{approx}) and again yields a lengthy expression, which we choose not to display here.

After the above integrations, we are again in the position to display the anomalous dimensions for the massive and massless fermions as functions of the energy scale $\mu$. (See Fig.~\ref{gamma_mu_p}.) For all choices of parameters, at high scales, where the masses of the massless particles are still comparatively small, the anomalous dimensions for both particle species coincide as they should. They are to be found in (pre freeze out) plateaus at the same value as in the fully massive case for the same value of the parameter $c$. Once the massive fermions start freezing out, their anomalous dimension is left behind by that of the massless fermions (dotted lines), which goes towards the critical value for the onset of chiral condensation. Usually, when the freeze out begins, the anomalous dimension of the massive fermions will (at least at first) follow the increase of the anomalous dimension of the massless quarks. Only when there are more massless than massive quarks (See, for example the case $n_f=3N_f$ in the middle panel.) the anomalous dimension of the massive fermions can also start decreasing immediately. If the critical anomalous dimension is reached sufficiently late in the freeze out process of the massive fermions, their anomalous dimension will begin falling again. If the freeze out is almost complete at the onset of chiral condensation, the anomalous dimension of the massive fermions goes to zero. This happens in those setups in which also a post freeze out plateau develops in the anomalous dimension of the anomalous dimension of the massless fermions. Very importantly, in the partially massive case post freeze out plateaus in the anomalous dimension of the massless flavours appear also for a critical value of the anomalous dimension for chiral condensation of 1. (See middle panel.) It is no longer a special feature of the case where this value equals 2. Further, there can also be a pre freeze out plateau at nonzero anomalous dimension. (See the black curves in the middle panel.) In this context, the larger the number of massless flavours as compared to the number of massive flavours, the larger is the value of the anomalous dimension in the pre freeze out plateau. This statement also holds in the case, where the anomalous dimension at the onset of chiral condensation equals 2.

\begin{figure*}[t]
\hfill\subfigure{\resizebox{!}{5.8cm}{\includegraphics{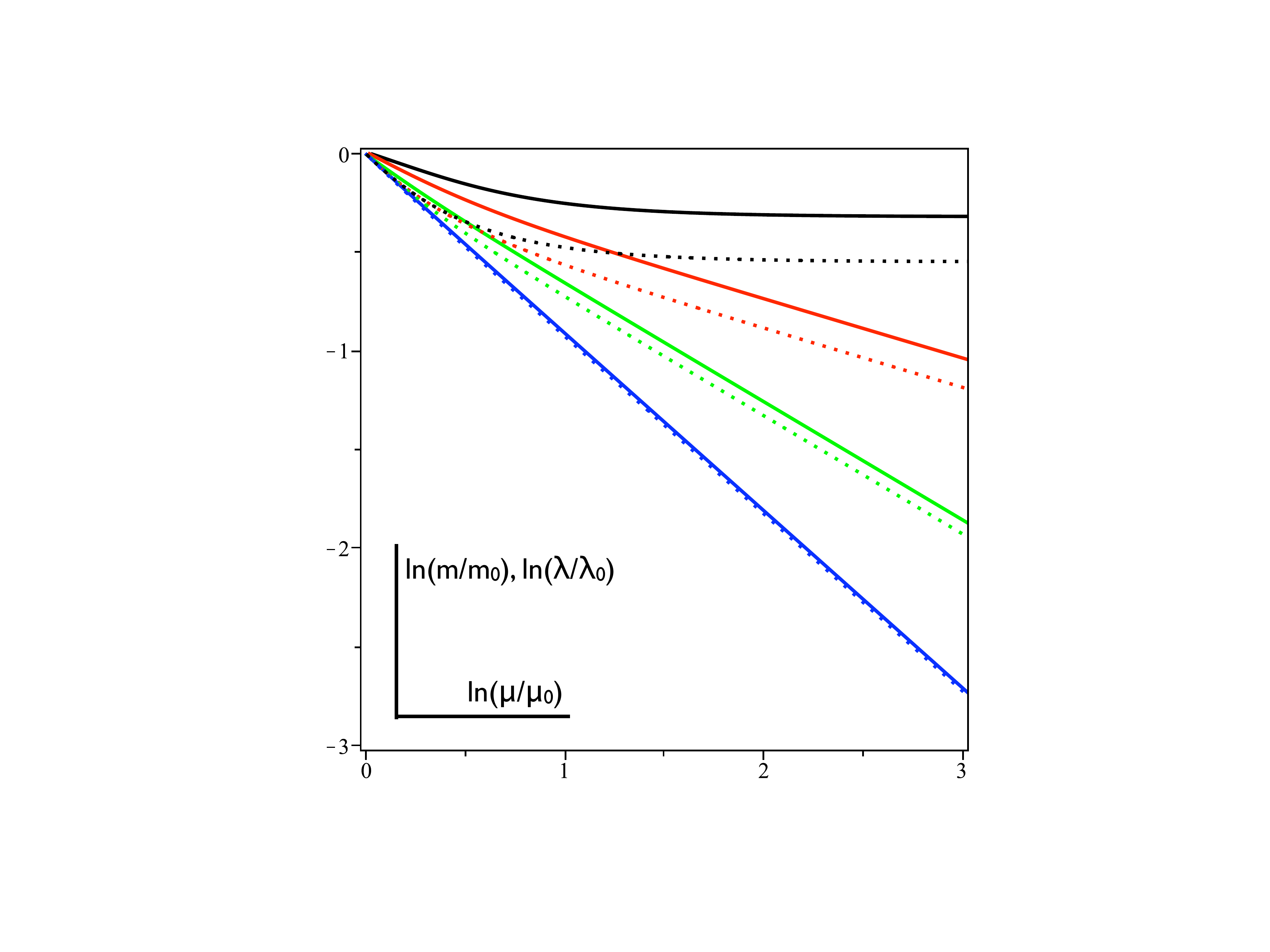}}}\hfill\subfigure{\resizebox{!}{5.8cm}{\includegraphics{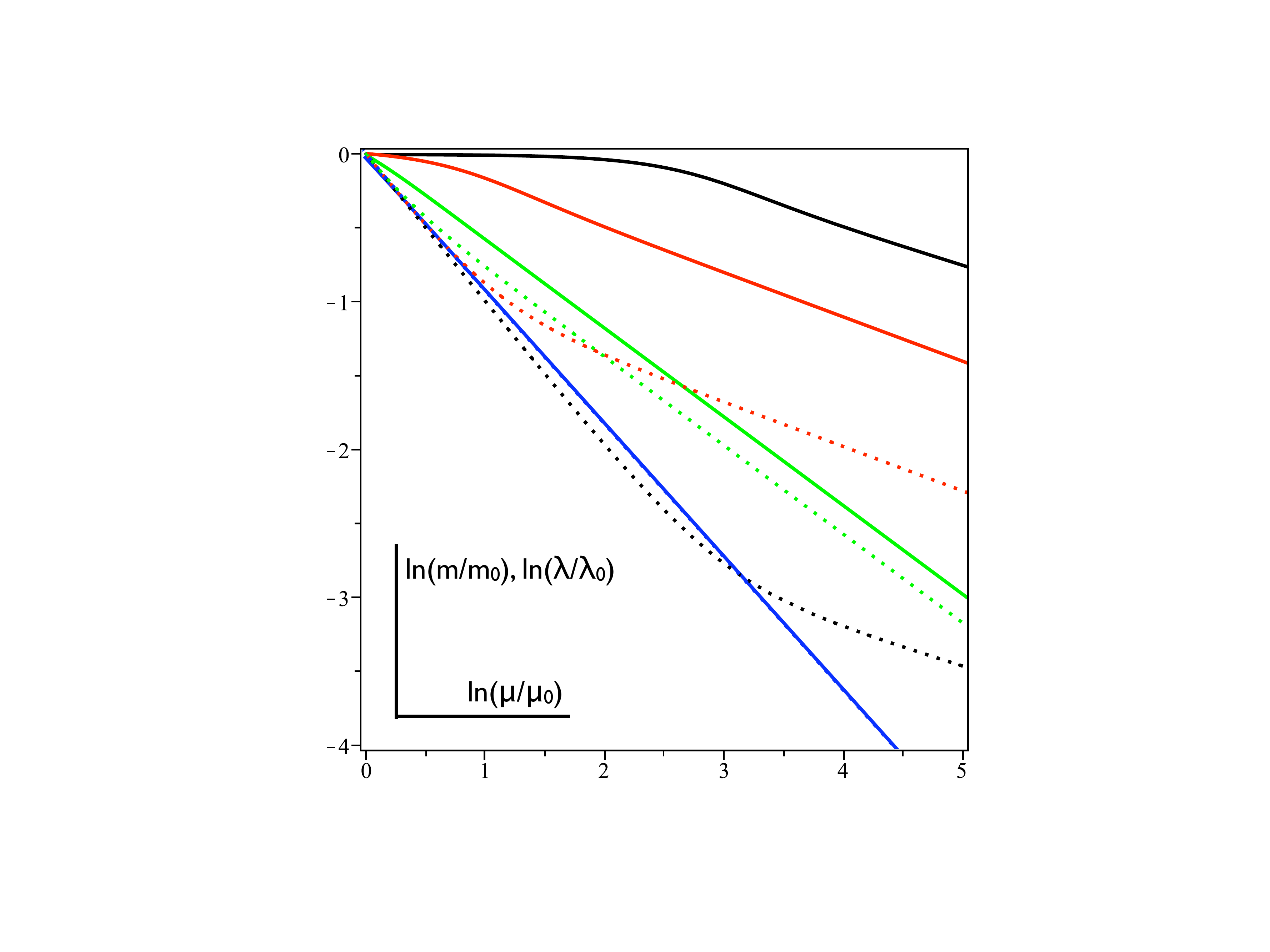}}}\hfill\subfigure{\resizebox{!}{5.8cm}{\includegraphics{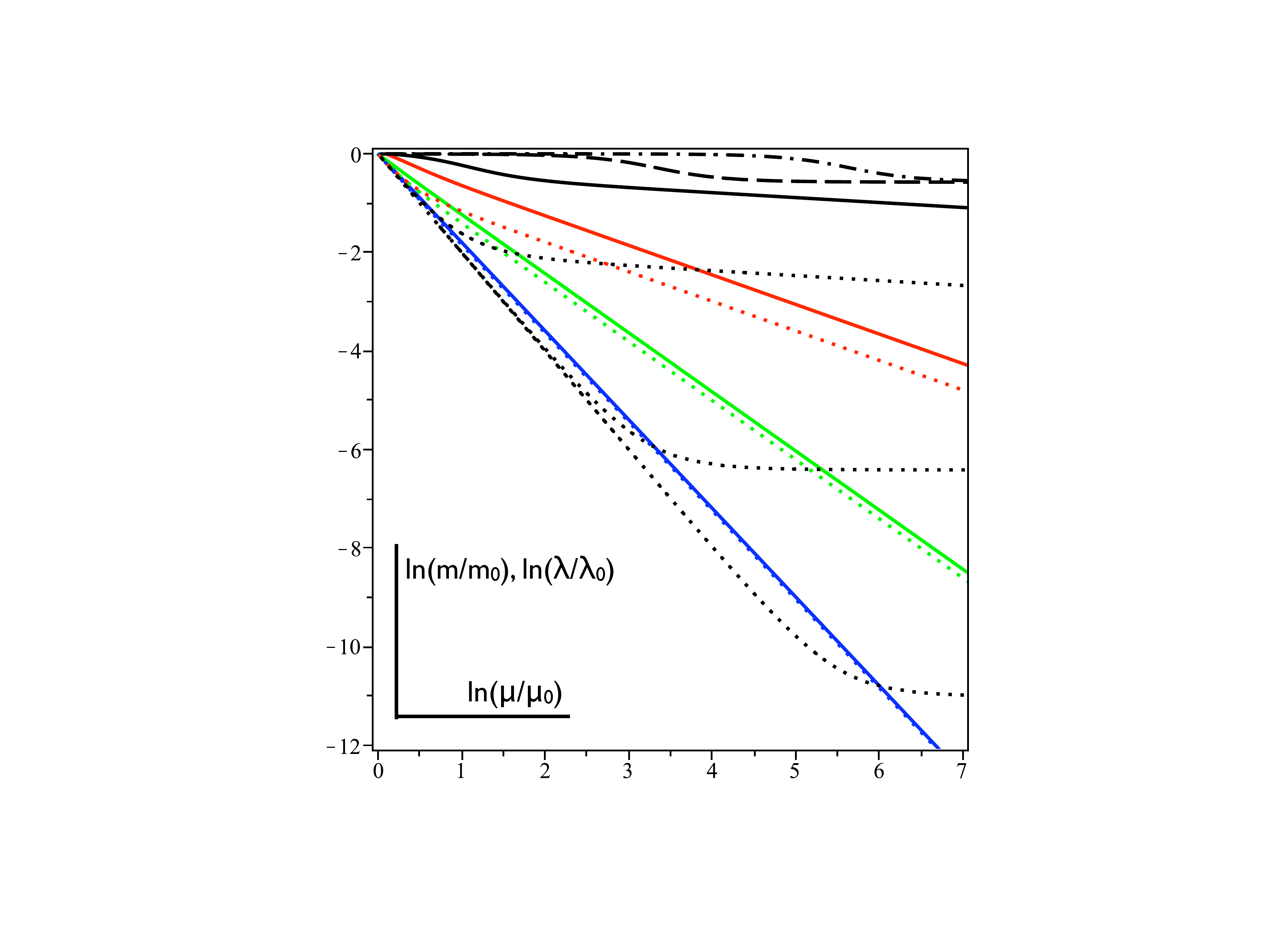}}}\hfill~
\caption{Renormalisation of the mass operators of the massless (dotted) and the massive (other) flavours. Left panel, $\gamma_c=1$, $N_f=n_f$; from bottom to top, $c=2.9$ (blue), $c=2.6$ (green), $c=2.3$ (red), $c=2.0$ (black). Middle panel, $\gamma_c=1$, $3N_f=n_f$; from top to bottom, $c=2.9$ (blue), $c=2.6$ (green), $c=2.3$ (red), $c=2.251$ (black). Right panel, $\gamma_c=2$, $N_f=n_f$; from bottom to top, solid: $c=3.8$ (blue), $c=3.2$ (green), $c=2.6$ (red), $c=2.1$ (black); not solid: $c=2.001$ (dashed), $c=2.00001$ (dash-dotted). 
}
\label{m_mu_p}
\end{figure*}

Figure \ref{m_mu_p} displays the logarithms of the renormalisation factors of the mass operators as a function of the logarithm of the scale prior to chiral condensation. The dotted lines are for the massless fermions, the others for the massive ones. (Nearly) horizontal stretches correspond to small anomalous dimensions and steep stretches to large values of the anomalous dimension. The post freeze out plateaus are characterised by a maximally steep approach to the origin of the renormalisation of the massless flavours. The achieved renormalisations generically exceed the renormalisation achieved with a pre freeze out plateau. If a pre freeze out plateau is situated at a large value of the anomalous dimension it approaches the origin at an almost maximal slope and bends into the maximal slope (maximal anomalous dimension) just before finally meeting the origin.

\begin{figure*}[t]
\hfill\subfigure{\resizebox{!}{5.8cm}{\includegraphics{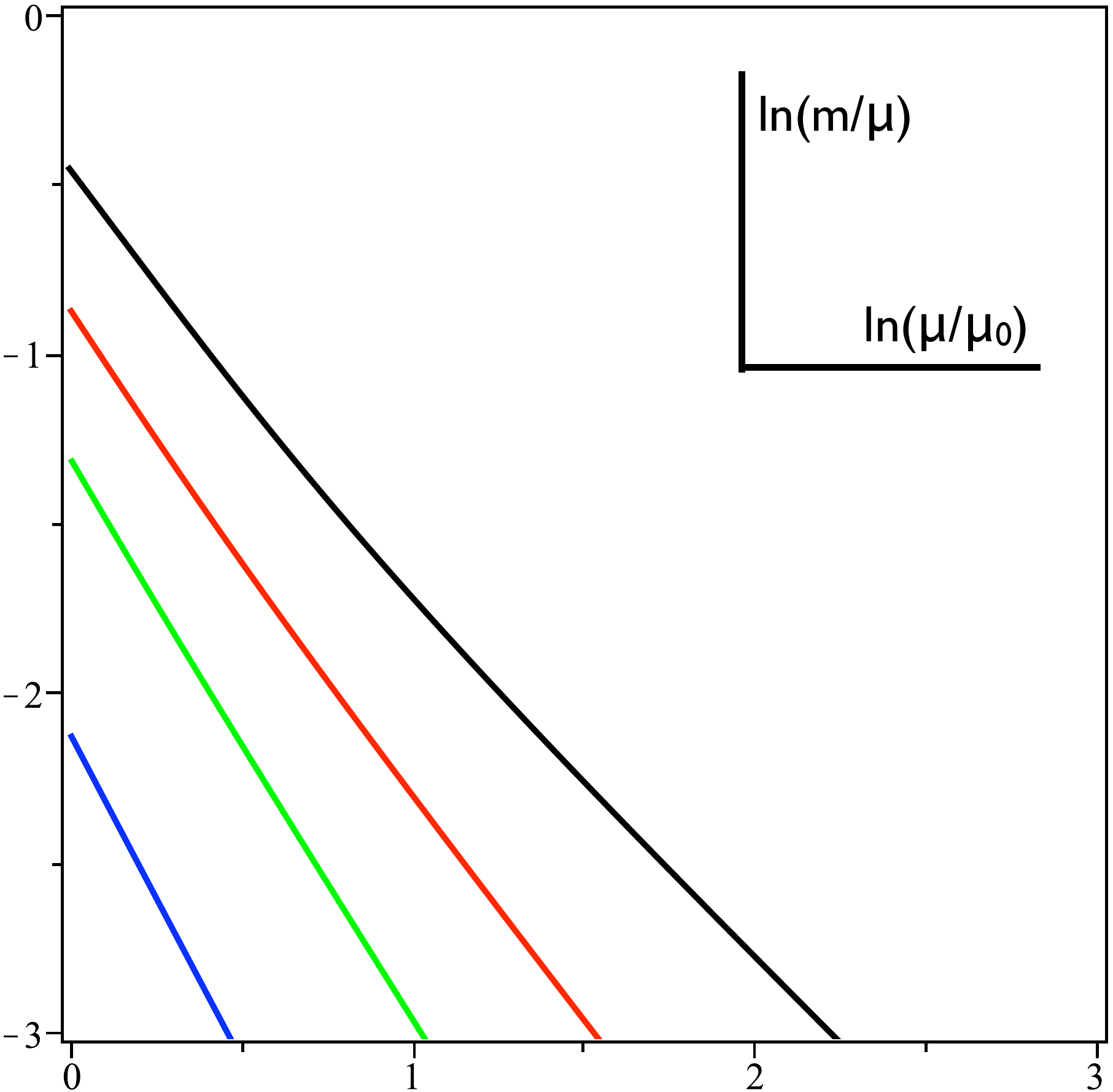}}}\hfill\subfigure{\resizebox{!}{5.8cm}{\includegraphics{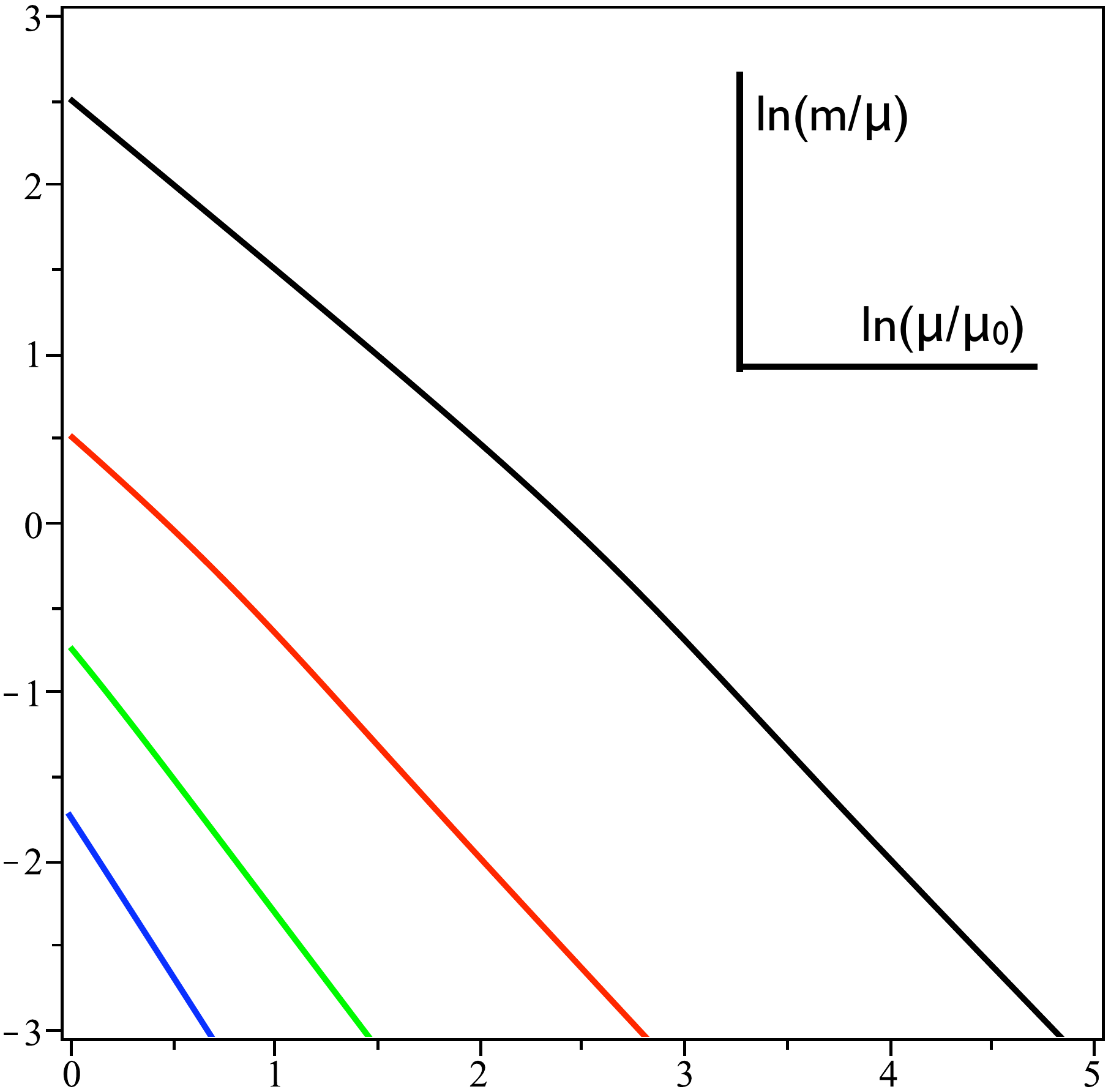}}}\hfill\subfigure{\resizebox{!}{5.8cm}{\includegraphics{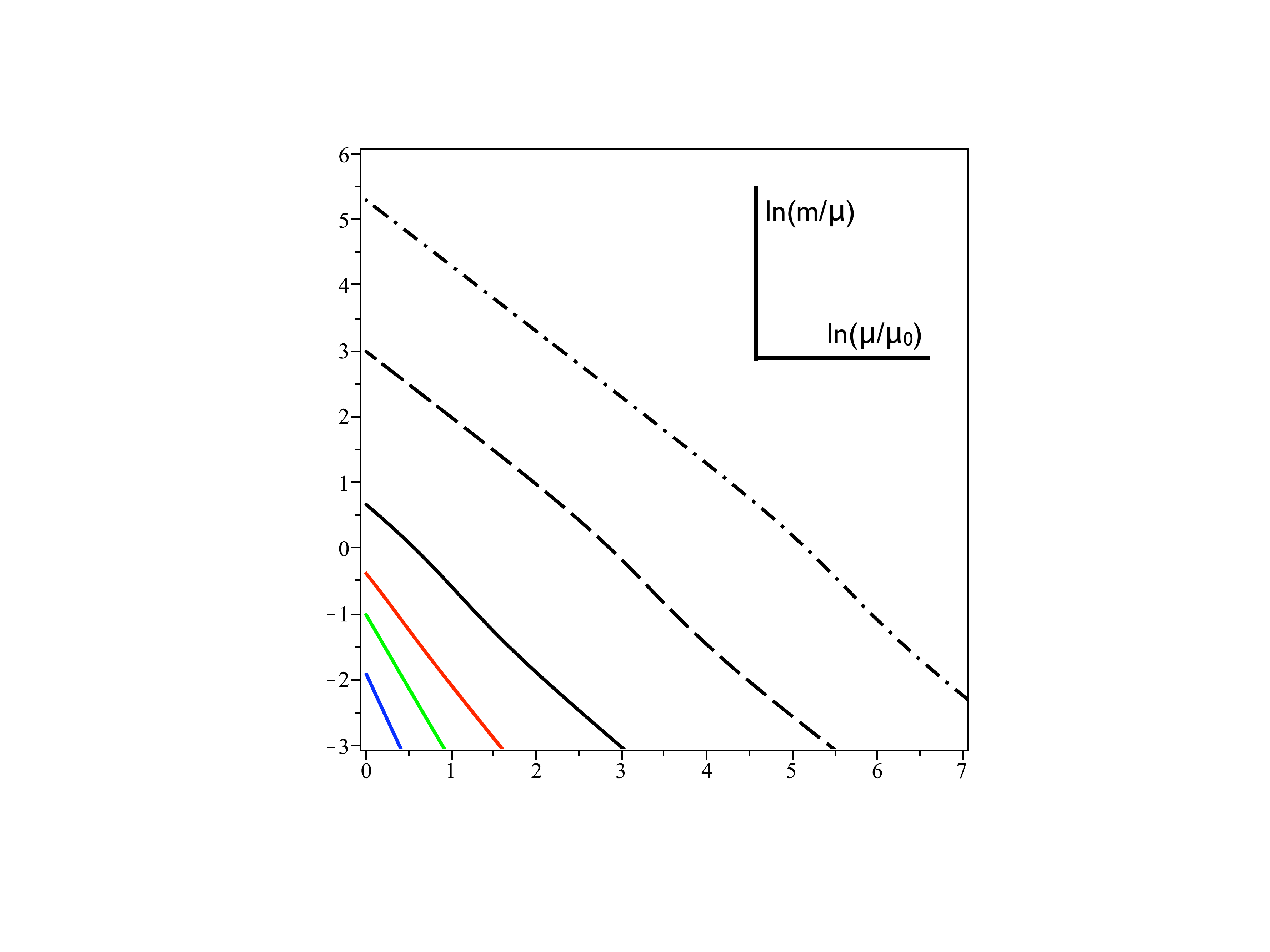}}}\hfill~
\caption{Logarithm of the mass to scale ratio as a function of the energy scale. Left panel, $\gamma_c=1$, $N_f=n_f$; from bottom to top, $c=2.9$ (blue), $c=2.6$ (green), $c=2.3$ (red), $c=2.0$ (black). Middle panel, $\gamma_c=2$, $3N_f=n_f$; from top to bottom, $c=2.9$ (blue), $c=2.6$ (green), $c=2.3$ (red), $c=2.251$ (black). Right panel $\gamma_c=2$, $N_f=n_f$; from bottom to top, solid: $c=3.8$ (blue), $c=3.2$ (green), $c=2.6$ (red), $c=2.1$ (black); not solid: $c=2.001$ (dashed), $c=2.00001$ (dash-dotted). 
}
\label{x_mu_p}
\end{figure*}

Figure \ref{x_mu_p} presents the logarithm of the mass to scale ration as function of the energy scale. Apart from the overall evolution of the ratio, the intercepts at the scale at which chiral condensation sets in are of particular interest. For situations with post freeze out plateau, these intercepts are especially large. That means that in these situations the renormalisation group enhancement of the condensate of the massless fermions takes place, while the extra massive quarks are much heavier than the scale. 
To the contrary, for models with pre freeze out plateau the intercepts are small; the massive quarks are lighter than the energy scale at the onset of chiral condensation. At the same moment their anomalous dimensions are large as well, which implies that they also might partake in the chiral condensation process. In the limit of light massive quarks we can resort to the Gell-Mann--Oakes--Renner relation to estimate the masses of the pions incorporating one (``light-heavy pion") or two (``doubly heavy pion") of the massive flavours:
$
m^2_{\pi_\mathrm{hl}}=1m\langle Q\bar{Q}\rangle f_\pi^{-2}
$
or
$
m^2_{\pi_\mathrm{hh}}=2m\langle Q\bar{Q}\rangle f_\pi^{-2} 
$
where, in a technicolour setting, $\langle Q\bar{Q}\rangle=O(\mathrm{TeV}^3)$ and $N_f^\mathrm{cond}f_\pi^2=2\Lambda_\mathrm{ew}^2$. $N_f^\mathrm{cond}$ is the number of flavours participating in the condensation process. From there we get $(m_{\pi_\mathrm{hl}}/\mathrm{TeV})^2=8N_f^\mathrm{cond}\times(m/\mathrm{TeV})$. For the smallest intercept in the right panel of Fig.~\ref{x_mu_p} of $(m/\mathrm{TeV})\lesssim-2$ this yields $m_{\pi_\mathrm{hl}}/\mathrm{TeV}\approx(N_f^\mathrm{cond})^{1/2}$. 

Deviations from the present analysis arise form the term 
$-\frac{3}{2}\frac{1}{T(R)}\frac{\delta}{(N_f+n_f/b_0)}$. Hence, we have to study
\begin{eqnarray}
\int_{\mu_1}^{\mu_2} \frac{d\mu}{\mu} \frac{\delta}{\tilde N_f}
=\hskip 6cm\\\nonumber=
\left[\frac{8\pi^2}{g^2}\langle\tilde N_f^{-1}\rangle+2\ln(g)C_2(G)\left\langle\frac{1+2\frac{\bar{\beta}^\prime_0}{\bar{\beta}_0}}{\tilde N_f}\right\rangle\right]_{g(\mu_1)}^{g(\mu_2)} ,
\end{eqnarray}
where $\tilde N_f=N_f+n_f/b_0$.
$\langle\dots\rangle$ again indicates an intermediate value in the sense of the mean value theorem. 
If the anomalous dimension changes slowly, the $\beta$-function remains close to zero, $g(\mu_2)$ stays close to $g(\mu_1)$, and the deviation from our result is small. Where the anomalous dimension changes at a higher rate the logarithmic term leads to an enhancement of the mass renormalisation and the $g^{-2}$ term to a reduction.
We can estimate once more,
\begin{eqnarray}
\int_{\ln\mu_1}^{\ln\mu_2} d\ln\mu ~ \frac{\delta}{\tilde N_f}
<
\langle\tilde N_F^{-1}\rangle\frac{8\pi^2}{[g(\mu_2)]^2}
=\nonumber\\=
\langle\tilde N_F^{-1}\rangle3\frac{C_2(R)}{\hat{\gamma}_{1\bigcirc}(\mu_2)}
<
\langle\tilde N_F^{-1}\rangle3\frac{C_2(R)}{\hat{\gamma}(\mu_2)} .
\end{eqnarray}
For an analysis in the range of large masses, $\langle\tilde N_F^{-1}\rangle\approx\langle b_0/n_f\rangle$. As there $b_0\ll1$, large masses further reduce the deviation. As in the previous fully massive case this shows that the behaviour in the vicinity of the post freeze out plateau is well captured in the present approximation. Around the pre freeze out plateau the accuracy is again assured while the value of $\int(1-b_0)d\ln\mu$ as integrated backward from the plateau towards smaller values of the scale remains small compared to unity.

In areas where $b_0$ is changing, that is, for finite $\delta$, the effective value of $C$ is reduced. 
For the curves depicting the anomalous dimensions for the massless flavours in the left panel of Fig.~\ref{gamma_mu_p} this will lead to a steepening in the freeze out region, as seen before in the fully massive case. At the larger values of the parameter $C$ this also holds for the anomalous dimensions of the massive particles. At smaller values the flattening out and, ultimately, the relative maximum seen in the black curve could become visible due to the shift in the effective value of $C$.
For the middle panel the steepened increase of the anomalous dimension of the massless fermions persists for the larger values of $C$. There, the anomalous dimension of the massive flavours is falling and the rate of this falloff is also increased by an effective reduction of $C$. For smaller values of the parameter $C$ an effective reduction tends to introduce a small local maximum and afterwards a falloff down to zero into the curves for the anomalous dimension of the massive fermions. In the curve for the massless flavours a post freeze out plateau is introduced and stabilised. The slope connecting the pre to the post freeze out plateau is universal and should not be affected.
What has been said here about the curves in the middle panel holds also for the right panel: For large values of $C$ an effective reduction of $C$ in the freeze out region steepens the approach of the anomalous dimension of the massless fermions to the critical value of the anomalous dimension; for smaller values the post freeze out plateau is stabilised. For the massive flavours first a maximum is introduced (at larger $C$) and then a drop down to vanishing anomalous dimensions, while the maximum (seen at smaller $C$) is pushed to the left.

\subsubsection{Outlook I}

One possible generalisation of what has been discussed here so far could address systems with more than two different types of flavours characterised by their mass. The lightest flavours will always set the boundary condition because their anomalous dimension will always reach the critical value first. The fermions for which the number of active flavours changes most in the interesting range of scales would influence the dynamics most.

Due to what is perceived as a relatively small matter content, one commonly does not expect quantum chromodynamics to be close to a fixed point in the renormalisation group evolution when chiral symmetry is breaking. Hence, the present analysis, which is based on the nearness of a would-be fixed point should receive important corrections. One does not expect any plateaus, neither pre nor post freeze out. The pattern of two light flavours ($u$ and $d$) in conjunction with one flavour ($s$) that is actively freezing out at the scale where chiral symmetry breaks is, however, reminiscent of what has been discussed above. (Actually, according to the description of the effective number of active flavours the $c$ and even the $b$ quark are not fully frozen out, while the $t$ quark is to a better approximation infinitely massive than the $u$ and $d$ are massless.) As a consequence, if the dynamics of chiral symmetry breaking in quantum chromodynamics should be dominated by the freeze out of the $s$ quark, the universal slope in the freeze out curves might play a role even there, in the sense, that the final approach of the anomalous dimension of the $u$ and the $d$ quark looks like the upper end of the black dotted curve in the right panel of Fig.~\ref{gamma_mu_p} and possibly continues steeper down to perturbatively small values. (The $s$ quark could not play such a potential important role in this context if it were not provided with a fitting explicit mass by the breaking of the electroweak symmetry.)

Another extension could deal with fermions with different representations on top of different masses, as the massive all-order $\beta$-function (\ref{maob}) also allows for fermions in different representations to be present simultaneously. This requires information on the relationship between the anomalous dimensions of the flavours transforming under different representations. 


\subsubsection{Outlook II: enhanced flavour symmetry}

For a moment, let us consider technicolour models with two flavours. 
(Concrete models are realised in the form of minimal and next-to-minimal walking technicolour \cite{Sannino:2004qp,Dietrich:2005jn,Dietrich:2005wk}.)

Such theories with techniquarks in non-(pseudo)real representation of the technicolour gauge group feature an $SU(2)_L\times SU(2)_R\rightarrow SU(2)_V$ chiral symmetry breaking pattern. It entails always the correct breaking of the electroweak symmetry, $SU(2)_L\times U(1)_Y\rightarrow U(1)_\mathrm{em}$ \cite{Peskin:1980gc}. 

For techniquarks in a strictly real representation the breaking pattern is $SU(4)\rightarrow SO(4)$.
Electroweak radiative correction provide the 6 Nambu--Goldstone bosons that do not become the longitudinal degrees of freedom of the weak gauge bosons with positive squared masses, which stabilises the correct vacuum alignment \cite{Peskin:1980gc} and takes them beyond the direct exclusion limit for technipions \cite{Dietrich:2009ix,Amsler:2008zzb}.

For a pseudoreal representation the breaking pattern is $SU(4)\rightarrow Sp(4)$ with two extra technipions, which receive negative contributions to their squared masses from electroweak radiative corrections, which, in turn, destabilises the correct embedding. This has to be counteracted by an appropriate mechanism (extended technicolour).

All it takes to break the electroweak symmetry appropriately is an $SU(2)_L\times SU(2)_R\rightarrow SU(2)_V$ chiral symmetry breaking pattern. Hence, above the technicolour scale not the full $SU(4)$ symmetry needs to be preserved, but merely the $SU(2)_L\times SU(2)_R$ subgroup:
The pure two-flavour technicolour sector is made up of 4 Weyl fermions, which can be collected in a column vector
\[
\left(
\begin{array}{r}
U_L\\D_L\\-i\sigma^2U_R^*\\-i\sigma^2D_R^*
\end{array}
\right)
\]
in which all components transform as left-fields. This makes the $SU(4)$ flavour symmetry more obvious. What is called left and what right is determined by the way the electroweak sector is coupled in. For the flavour symmetry the mass terms
\begin{equation}
\mathcal{L}_m = m~(\bar{U}_LU_R + \bar{U}_RU_L + \bar{D}_LD_R + \bar{D}_RD_L)
\end{equation}
and
\begin{equation}
\mathcal{L}_\lambda= \lambda~(\bar{U}_LD_L + \bar{D}_LU_L + \bar{U}_RD_R + \bar{D}_RU_R)
\end{equation}
are equivalent. They can be expressed by contracting two of the above column vectors with the matrices
\begin{equation}
\left(\begin{array}{cc}\mathbbm{O}&\mathbbm{1}\\\mathbbm{1}&\mathbbm{O}\end{array}\right)
\mathrm{~~~or~~~}
\left(\begin{array}{cc}\mathbbm{1}&\mathbbm{O}\\\mathbbm{O}&-\mathbbm{1}\end{array}\right) .
\end{equation}
$\mathbbm{O}$ and $\mathbbm{1}$ stand for $2\times2$ zero and unit matrices, respectively.

Only the first mass term breaks the electroweak symmetry. The second leaves a residual $SO(4)\simeq SU(2)_L\times SU(2)_R$ flavour symmetry behind.
It contributes to the masses of the Nambu--Goldstone bosons that correspond to generators that link left with right fields. (These are the modes with finite technibaryon number.) 
For techniquarks in a pseudoreal representation of the technicolour gauge group, terms which break the $SU(4)$ flavour symmetry to $SU(2)_L\times SU(2)_R$ are needed to stabilise the vacuum.
Another motivation for studying the interplay between an explicit $\mathcal{L}_\lambda$ mass term with a chiral condensation taking place in the $\mathcal{L}_m$ channel is to regulate the amount of walking of the theory or to circumvent the exact conformality of a given setup \cite{Hietanen:2008mr}.
As opposed to the partially massive case analysed in detail above, where the massive and massless fields are clearly separated, the last mentioned case with $\mathcal{L}_\lambda$ and $\mathcal{L}_m$ channels does not allow for such a clear separation of the fields. Therefore, a corresponding treatment in the present framework requires further study. 


\subsection{Use in constructing walking technicolour models\label{USE}}

Walking dynamics in technicolour theories serve to relieve the tension between the large mass of the top quark, which has to be generated by extended technicolour interactions, and the small experimental bounds on flavour changing neutral currents, by a sizeable renormalisation of the techniquark condensate. With all flavours exactly massless these models are to be found just below the conformal window in the $N_c$-$N_f$ plane.
Here the approach is slightly different from massless walking technicolour, as we are considering theories that would be inside the conformal window and would thus evolve into an infrared fixed point, if all their techniquarks were massless. We regard instead a theory where at least two flavours are massless and gauged under the electroweak. This assures that the electroweak symmetry is not broken explicitly before the chiral symmetry is broken spontaneously. (Already in theories with all fermion massless and more than two flavours it is advantageous to gauge only two of them: Doing so makes it easier to avoid experimental bounds on the oblique parameters and solves the vacuum alignment problem, if the representation of the two gauged flavours does not happen to be pseudoreal \cite{Peskin:1980gc}. These are the so-called partially gauged technicolour models introduced in \cite{Dietrich:2005jn}.) In order to have an efficient renormalisation enhancement of the techniquark condensate the anomalous dimension of the massless flavours has to stay close to a large value---optimally, close to its critical value---for a sizeable range of energy scales before chiral condensation sets in. The above analysis indicates that this can be achieved in two different ways in the present context, by exploiting the pre freeze out plateau or the post freeze out plateau.
{In theories with pre freeze out plateau, the mass of the massive fermions is small compared to the energy scale.}
{In theories with post freeze out plateau, the mass of the massive fermions is large compared to the energy scale.}
{Hence, in the latter case, we have walking at maximal anomalous dimension with heavy massive fermions as opposed to walking at almost maximal anomalous dimension with light massive fermions.}

\begin{figure*}[t]
\hfill\subfigure{\resizebox{!}{5.8cm}{\includegraphics{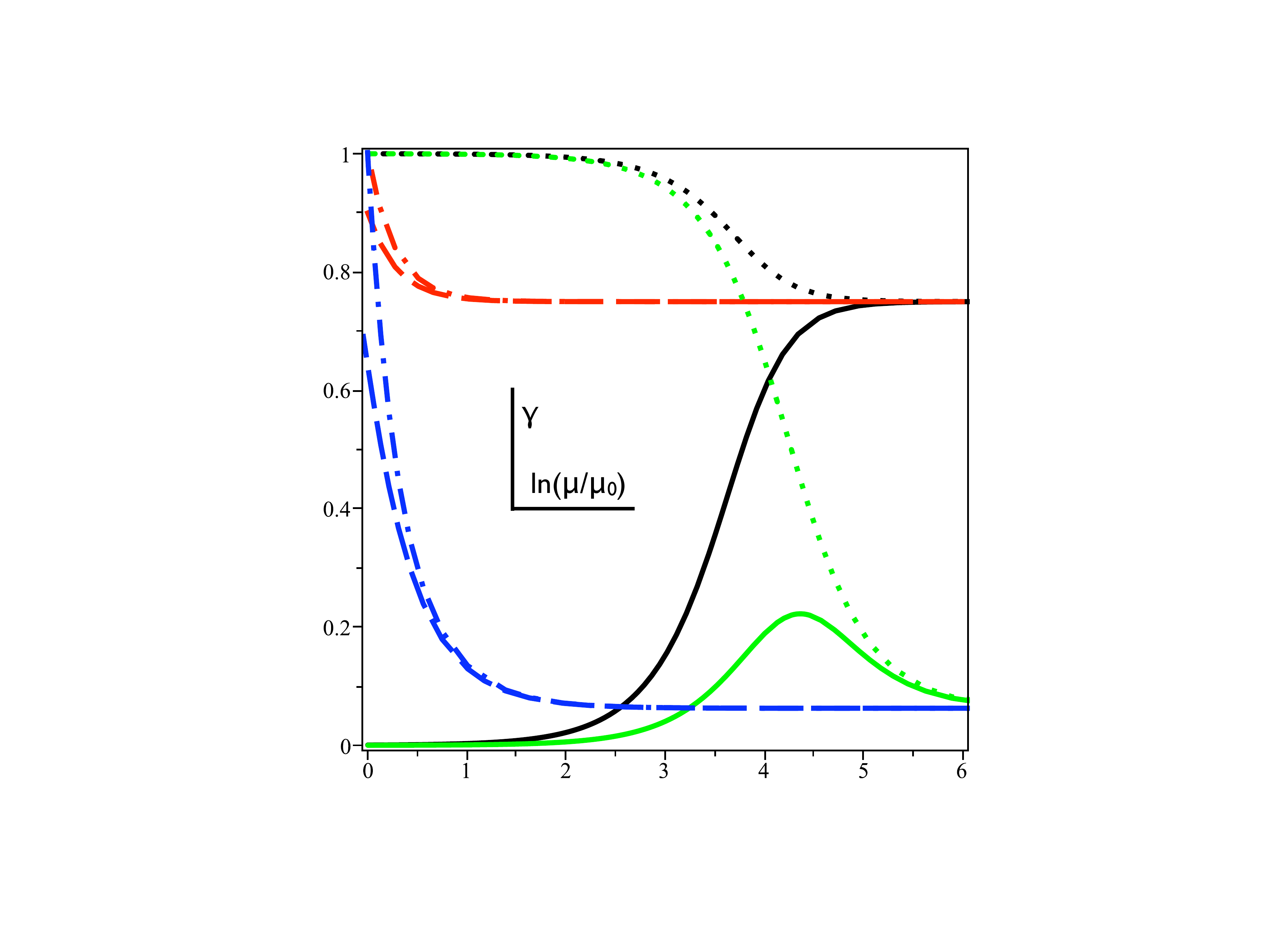}}}\hfill\subfigure{\resizebox{!}{5.8cm}{\includegraphics{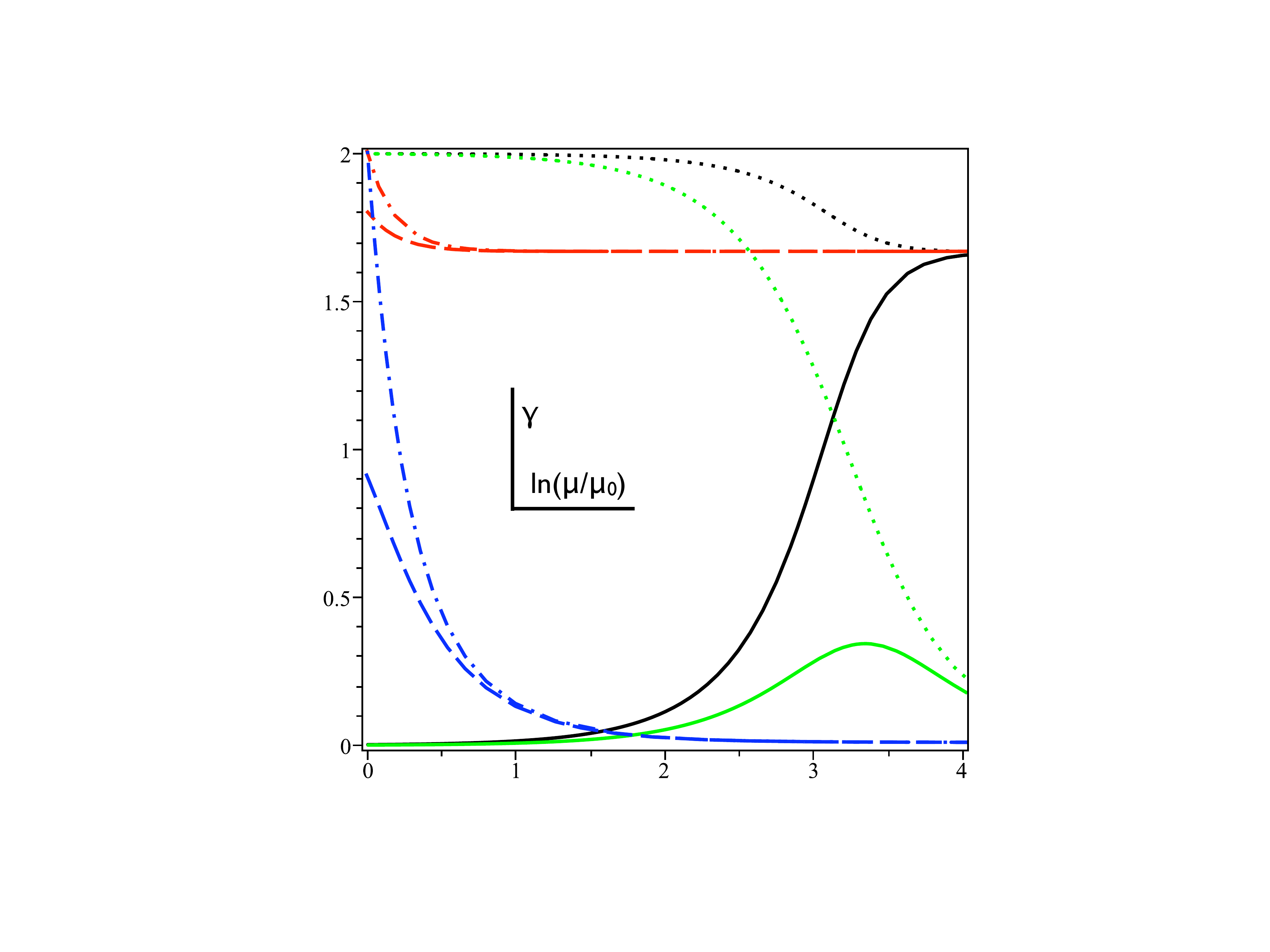}}}\hfill~
\caption{Anomalous dimension as a function of the energy scale for different setups within the same technicolour gauge group. Left panel, $\gamma_c=1$, gauge group $SU(3)$: black and red $N_f+n_f=12$ (just inside conformal window); black $N_f=1$ (minimal number of massive flavours), solid anomalous dimension of the massive flavour, dashed anomalous dimension of the massless flavours; red $n_f=2$ (minimal number of massless flavours), dashed anomalous dimension of massive flavours, dash-dotted anomalous dimension of the massless flavours; green and blue $N_f+n_f=16$ (just asymptotically free); green $n_f=11$ (maximal number of massless flavours), solid anomalous dimension of the massive flavours, dotted anomalous dimension of the massless flavours; blue $n_f=2$ (minimal number of massless flavours), dashed anomalous dimension of the massive flavours, dash-dotted anomalous dimension of the massless flavours. Right panel, $\gamma_c=2$, gauge group $SU(4)$: everything exactly the same as for the left panel, only for green and blue $N_f+n_f=22$ (just asymptotically free)
In both cases the length of the post freeze out plateaus (if present) has been shortened below the actual length to allow a representation in a single plot by detuning the parameters away from the actual values.
}
\label{gamma_mu_su}
\end{figure*}

Massless walking technicolour theories have their number of flavours outside the conformal window, but as close as possible to the lower bound of the latter.

\begin{table}[t]
\begin{tabular}{llrrrrrr}
R&$N_c$&$\pi S_\mathrm{pert}$&$N_f^{\gamma_c=2}$&$N_f^{\gamma_c=1}$&$N_f^\mathrm{a.f.}$&$\gamma_\mathrm{plateau}^{\gamma_c=2}$&$\gamma_\mathrm{plateau}^{\gamma_c=1}$\\\hline
F&2\footnote{F2 is a pseudoreal representation; the vacuum alignment has to be taken care of.}&0.3&5.5&7.3&11.0&1.7&0.8\\
F&3\footnote{A Witten anomaly has to be removed.}&0.5&8.3&11.0&16.5&1.7&0.8\\
F&4&0.7&11.0&14.7&22.0&1.7&0.9\\
F&5$^b$&0.8&13.8&18.3&27.5&1.9&0.9\\
F&6&1.0&16.5&22.0&33.0&1.9&0.9\\
G&2$^b$&0.3&1.4&1.8&2.8&\footnote{$N_f^{\gamma_c=\bullet}<2$}&$^c$\\
G&3&0.7&1.4&1.8&2.8&$^c$&$^c$\\
S\footnote{S2 coincides with G2.}&3&1.0&1.7&2.2&3.3&$^c$&0.2\\
A\footnote{A3 coincides with F3.}&4\footnote{Apart from $S_\mathrm{pert}$ values for A4 coincide with values for F2.}&1.0&5.5&7.3&11.0&1.7&0.8
\end{tabular}
\caption{List of all $SU(N)$ theories with a small perturbative $S$ parameter. ($\pi S_\mathrm{pert}\le1$ when only two flavours are gauged.) The two benchmark values, 1 and 2, for the critical value of the anomalous dimension at which chiral condensation sets in, are analysed. R stands for the representation (F=fundamental, G=adjoint, S=two-index symmetric, A=two-index antisymmetric), $N_f^{\gamma_c=2}$ ($N_f^{\gamma_c=1}$) for the lower bound of the conformal window if the critical value of the anomalous dimension for the onset of chiral condensation equals 2 (1), and $N_f^\mathrm{a.f.}$ for the total number of flavours above which asymptotic freedom is lost. The column marked by $\gamma_\mathrm{plateau}^{\gamma_c=2}$ ($\gamma_\mathrm{plateau}^{\gamma_c=1}$) shows the value of the anomalous dimension in the pre freeze out plateau for a total number of flavours that is just inside the would-be conformal window.}
\label{table}
\end{table}

In theories with a pre freeze out plateau the total number $n_f+N_f$ of flavours is just inside the would-be conformal window. This is because, for the renormalisation of the mass operator of the massless flavours to be as efficient as possible the plateau value of the anomalous dimension of the mass operator of the massless techniquarks must be as close as possible to the critical value for the onset of chiral condensation as possible, and this is achieved by being just inside the would-be conformal window. The number $n_f$ of massless fermions must be at least 2, to accommodate the dynamical electroweak symmetry breaking. Other than that, it is probably most advantageous to take all other flavours massive for reasons of vacuum alignment and direct discovery limits, for example, for extra Nambu-Goldstone modes. The presence of a pre freeze out plateau at a value of the anomalous dimension close to the critical is a generic feature in the sense that it is present for most gauge groups where the conformal window starts above 2 flavours (the minimal number we keep massless) and asymptotic freedom is not lost for 3 flavours (at least one additional massive flavour). This can be seen in the two last columns of Tab.~\ref{table}, which lists all theories based on $SU(N_c)$ gauge groups, which for two electroweakly gauged flavours have a sufficiently small contribution to the perturbative $S_\mathrm{pert}$ parameter. (Available electroweak precision data tells us that the $S$ parameter should be small. In walking theories its perturbative value is a conservative upper estimate for its value.) In most cases the anomalous dimension is close to the critical value and leads thus to an efficient renormalisation of the techniquark condensate.

In theories with a post freeze out plateau, the number $n_f$ of massless flavours must be just outside the would-be conformal window. The total number $n_f+N_f$ of techniquarks can, in principle, be as large as allowed by asymptotic freedom. 
A number of only $N_f=1$ massive flavours has the advantage of leading to a transition which is as smooth as possible, which extends the plateau. Further, this choice also features a pre freeze out plateau with optimal plateau value for the anomalous dimension for the massless techniquarks. Lastly, due to the thus most gradual transition, the here used approximations are under best possible control.
The appearance of theories with a post freeze out plateau is less generic than that of theories with pre freeze out plateau. It depends on how far the next integer number of flavours is below the edge of the conformal window. (This is a feature this phenomenon has in common with the massless technicolour theories \cite{Dietrich:2006cm}.) The phenomenon can, however, be seen without fine-tuning, but by simply sticking to our benchmark values for the critical anomalous dimension and looking at various possible setups. In order to see this, consider the $SU(N_c)$ theories depicted in Fig.~\ref{gamma_mu_su}. The left panel features theories based on an $SU(3)$ technicolour group analysed with the critical anomalous dimension equal to one. The red and black curves are for theories just inside the conformal window ($n_f+N_f=12$). For any split of this number, there is a pre freeze out plateau. For the minimal number of massive flavours ($N_f=1$) there is additionally a post freeze out plateau. In fact, for the plot the parameters were detuned slightly to allow the representation in a single panel, as for the exact choice of parameters, the post freeze out plateau becomes infinitely long. For the blue and green curves, the maximally allowed number of flavours allowed by asymptotic freedom ($N_f+n_f=16$) has been chosen. For all possible splits of this number between massive and massless flavours, there is a pre freeze out plateau, but only at a very small value of the anomalous dimension. For the maximally allowed number of massless flavours for which the theory is not exactly conformal ($n_f=11$) we see again a post freeze out plateau (which again has been shortened by detuning). We do not see such a plateau if we only retain the minimal number ($n_f=2$) of massless flavours that are required to construct a model of dynamical electroweak symmetry breaking. Analogous findings are displayed in the right panel for an $SU(4)$ model in the fundamental representation and analysed with the critical anomalous dimension set to 2: Small total numbers of flavours lead to a more useful pre freeze out plateau and large numbers of massless flavours make it more likely to find a post freeze out plateau.


\section{Summary\label{SUMM}}

Here we have analysed farther the implications from the mass-dependent all-order $\beta$-function derived in \cite{Dietrich:2009ns}. There it had already been used to determine the lower bound of the quasiconformal window. Here we proceed to identify universal behaviour of the renormalisation group evolution in the vicinity of a would-be fixed point and to shed more light on what is to be understood under the quasiconformal window. (``Would-be" because the evolution would hit the fixed point in the absence of masses.) We analyse the cases where all fermions are massive and, phenomenologically more relevant, where part of them are massless prior to spontaneous chiral symmetry breaking. (In the partially massive case we rely on the assumption that the anomalous dimension for the fermion mass operator at fixed coupling and energy scale is a decreasing function of the fermion mass. In all cases we took the anomalous dimension as an increasing function of the coupling.) In both cases, we identify two kinds of walking behaviour in the anomalous dimension as a function of the energy scale. In one case the plateau is ended (lowest energy scale) by the onset of the freeze out of the massive fermions. The critical value of the anomalous dimension for the onset of chiral symmetry breaking is reached in the first half of the freeze out process. Where the plateau starts (highest energy scale) depends on the initial condition for the renormalisation group evolution. In the other case, the plateau begins when the massive flavours cannot freeze out much more, and it ends when the critical value of the anomalous dimension is finally reached. (The technical reason, why this second type of plateau can arise also when all flavours have the same mass is, in short, that in the course of the analysis the mass has to be taken to infinity for fixed anomalous dimension and not for fixed coupling.) For a pre freeze out plateau the critical anomalous dimension is reached when the massive flavours are much lighter than the energy scale. Then also their anomalous dimension is not much below the value of the massless fermions and all flavours may participate in chiral condensation. For a post freeze out plateau the massive flavours are much heavier than the energy scale. Both plateaux can be present simultaneously and they can appear for any value of the critical anomalous dimension. (This is reminiscent of the situation identified for quantum chromodynamics in \cite{Brodsky:2010vh}.) 

In walking technicolour theories the tension between the generation of the large mass of the top quark by extended technicolour (ETC) interactions and the measured smallness of flavour changing neutral currents is alleviated by a renormalisation enhancement of the techniquark condensate. In order for this process to be effective, the anomalous dimension of the mass operator of the techniquarks contributing to the chiral condensate has to be large (as close as possible to the critical value) for a sizeable range of scales.
The usual massless walking technicolour is constructed from massless techniquarks. Their total number must be as close as possible to the lower bound of the conformal window but outside. When allowing also for massive quarks the total number can also be inside the conformal window, only asymptotic freedom should of course not be lost. This opens the aisle for the construction of (partially) massive walking technicolour models where a freeze out of one or several massive techniquarks in the vicinity of a would-be fixed point provides for the walking behaviour. In this context it shows that such technicolour models with pre freeze out plateau are rather generic in the sense that as long as the choice of gauge group and representation allows us to accommodate at least two massless techniquarks below the conformal window and at least a third massive techniquark without losing asymptotic freedom, we find mostly at least one combination of massive and massless fermions which lets the theory walk in a pre freeze out plateau not too far below the critical value of the anomalous dimension. (See Tab.~\ref{table}.) Theories with a post freeze out plateau are less generic, but can be found without fine tuning. There the amount of walking depends on how far the next integer number of flavours is away from the critical number of flavours delimiting the conformal window. This is a feature they share with massless walking technicolour models. 

Given the persisting uncertainty in the determination of the lower bound of the conformal window, the present findings also allow us to more smoothly interpolate between models. Before, given the critical value for the anomalous dimension, a model was either an appropriate candidate for a viable walking technicolour theory or it did not have the required features. Now, in the same situation, the value of $\gamma_c$ sets limits on the masses of the massive fermions. For an especially striking example, compare the setups presented in green and blue, respectively in the right panel of Fig.~\ref{gamma_mu_su}. They have the same total number of flavours and as a consequence the same pre freeze out plateau value for the anomalous dimension, which happens to be too small for building a viable walking technicolour theory out of this setup. The green splitting between massive and massless flavours, however, features also a post freeze out plateau, which makes it a candidate theory. The extrapolation from the pre freeze out plateau would have ruled out both combinations as viable candidates. 

In this paper we have studied models with two types of techniquarks, one set of massless ones and another of mass degenerate massive ones. Another extension involves techniquarks in different simultaneously present representations, which can also be treated in the framework of the massive all-order $\beta$-function \cite{Dietrich:2009ns}. Yet another extension would be to extend the construction of models to allow for Weyl flavours in places and not only Dirac flavours. An additional motivation for the present and the previous study was also to get an idea of effects, which are caused by the fact that realistically we are not dealing with idealised technicolour, but with technicolour, where the flavour symmetry is broken by the coupling to the electroweak sector or by extended technicolour, which either has to ascertain the correct vacuum alignment, that extra technipions escape direct detection or, which probably leads to the strongest correction, that they are heavy enough to be viable candidates for cold dark matter.
It might even be extended technicolour effects that make a theory quasiconformal, which from its bare technicolour structure would be completely conformal and hence, not suited for breaking the electroweak symmetry dynamically.


\section*{Acknowledgments}

The author would like to thank
A.~Armoni,
R.~Barbieri,
S.~J.~Brodsky,
M.~J\"arvinen,
F.~Sannino,
K.~Socha,
and
A.~I.~Vainshtein
for discussions.
The work of DDD was supported by the Danish Natural Science Research Council.


\end{document}